\documentstyle[12pt]{article}
\newcommand{\gsim}{\mbox{ $
{}_{{}_{\textstyle\sim}}  \! \! \! \! \! {}^{{}_{\textstyle>}}$}}
\newcommand{\lsim}{\mbox{ $
{}_{{}_{\textstyle\sim}}  \! \! \! \! \! {}^{{}_{\textstyle<}}$}}
\begin{document}
\renewcommand{\thefootnote}{\fnsymbol{footnote}}
\thispagestyle{empty}
\vspace*{-1 cm}
\hspace*{\fill}  \mbox{WUE-ITP-95-021} \\
\vspace*{1 cm}
\begin{center}
{\large \bf Production and Decay of Neutralinos in
the Next-To-Minimal Supersymmetric Standard Model
\\ [3 ex] }
{\large F.~Franke\footnote{
email: fabian@physik.uni-wuerzburg.de},
H. Fraas
\\ [2 ex]
Institut f\"ur Theoretische Physik, Universit\"at W\"urzburg \\
D-97074 W\"urzburg, Germany}
\vfill

{\bf Abstract}
\end{center}

\vspace{1ex}

\noindent Within the framework of the
Next-To-Minimal Supersymmetric Standard Model
(NMSSM) we study neutralino production
$e^+e^- \longrightarrow \tilde{\chi}^0_i \tilde{\chi}^0_j$ ($i,j=1,\ldots ,5$)
at center-of-mass energies between 100 and 600 GeV
and the decays of the heavier neutralinos into the LSP plus
a fermion pair, a photon or a Higgs boson. For representative
gaugino/higgsino mixing scenarios, where the light neutralinos have
significant singlet components, we find some striking differences
between the NMSSM and the minimal supersymmetric model.
Since in the NMSSM neutralino and Higgs sector are strongly correlated, the
decay of the second lightest neutralino into a Higgs boson and the LSP
often is kinematically possible and even dominant in a
large parameter region of typical NMSSM scenarios.
Also, the decay rates into final states with a photon may be enhanced.

\vfill
\begin{center}
November 1995
\end{center}
\newpage
\setcounter{page}{1}
\section{Introduction}
Supersymmetry (SUSY) \cite{susy} may provide a solution to the hierarchy and
fine tuning problem of the standard model (SM) at the prize of more
than doubling the particle spectrum \cite{nilles}.
Therefore the search for supersymmetric particles is
one of the most challenging tasks at the present and future
high energy colliders. The most popular supersymmetric extension of the
SM is the Minimal Supersymmetric Standard Model (MSSM) \cite{haka}
characterized by a minimal particle content and a minimal number of
allowed couplings.
In order to give mass to both up and down quarks and
to avoid anomalies by higgsino loops, it contains two Higgs doublet fields
$H_1$ and $H_2$
with hypercharge $\pm 1/2$ and vacuum expectation values $v_1$ and $v_2$
($\tan\beta=v_2/v_1$).
An essential feature of the MSSM is the conservation of a new
quantum number called $R$-parity \cite{rpari} that implies two important
consequences: Supersymmetric particles can be produced only in pairs,
and the lightest supersymmetric particle (LSP) is stable.

One of the most promising processes to detect a supersymmetric signature
in $e^+e^-$ collisions is the pair production of neutralinos,
fermionic mass eigenstates composed of
the supersymmetric partners of photon, $Z$ boson and the neutral Higgs
bosons. Since cosmological reasons suggest the
lightest neutralino to be the LSP, at least one heavier neutralino must
be produced which can be identified by its subsequent decays.

In the MSSM, LEP data imply a lower bound of
23 GeV for the lightest neutralino \cite{neubounds}.
Neutralino production and decay has
been discussed in great detail in refs.~\cite{bartlneuprod, ambrosanio}
within the framework of the MSSM.

Many GUT and superstring theories \cite{barr,nilsredwy,derendinger},
however, favor the minimal extension of
the MSSM by a gauge singlet $N$ with hypercharge 0, the {\em Next-To-Minimal
Supersymmetric Standard Model} (NMSSM) \cite{drees, ellis}.
The most general superpotential of a supersymmetric model with an extra
gauge singlet is
\begin{equation}
\label{wsinglet}
W=\lambda H_1H_2N - \mu H_1H_2 - \frac{1}{3}kN^3 + \frac{1}{2} \mu' N^2
+\mu'' N,
\end{equation}
which reduces to that of the MSSM if $N$ is removed.
In the NMSSM one considers only the trilinear terms
\begin{equation}
\label{wnmssm}
W_{\mbox{\scriptsize NMSSM}}= \lambda H_1H_2N - \frac{1}{3}kN^3,
\end{equation}
so that the $\mu$ problem of the MSSM \cite{mu} is evaded.
Recently it was claimed that cosmological implications of the
NMSSM, namely the formation of domain walls at an early stage of the
universe, require the $Z_3$ symmetry of the superpotential to be
explicitly broken \cite{domain}. A solution to this domain wall problem
would be the reintroduction of the $\mu$ term in the superpotential.
Since an additional small $\mu$ term does not significantly affect the
masses and mixings of the neutralinos but increases the number of
free parameters we restrict ourselves to the superpotential of
eq.~(\ref{wnmssm}).

Thus the basic difference between the MSSM and the NMSSM or other
models with gauge singlets arise by the singlet components of neutralinos
and Higgs bosons which do not couple to fermions, gauge bosons and their
respective supersymmetric partners. In the NMSSM there are five
neutralinos instead of four in the MSSM. Also the neutral Higgs sector
is enlarged by one scalar and pseudoscalar Higgs bosons to three
CP even and two CP odd Higgs particles. As it was shown in refs.~\cite{
franke1, franke2}, experimental data still allows for massless
NMSSM neutralinos and Higgs bosons. We point out that these results
do not change with the reintroduction of an additional $\mu$ term in the
model, so that our choice of the superpotential eq.~(\ref{wnmssm})
instead of eq.~(\ref{wsinglet}) is well justified.

A further motivation for the NMSSM is the evasion of the usual
MSSM Higgs mass bounds \cite{quiros}. After including radiative
corrections, the theoretical upper bound
for the lightest Higgs scalar is increased by some 10 GeV compared to the
MSSM \cite{rad}.

Contrary to the minimal model, neutralino and Higgs sectors are strongly
correlated in the NMSSM. The masses and mixings of neutralinos are given
by the eigenvalues and eigenvectors of a $5\times 5$ matrix that depends
on the $SU(2)$ and $U(1)$ gaugino mass parameters $M$ and $M'$, the
singlet vacuum expectation value $x$, the ratio of the doublet
expectation values $\tan\beta$ and the couplings $\lambda$ and $k$ in the
superpotential. The properties of the neutral scalar and pseudoscalar
Higgs bosons follow from two $3 \times 3$ matrices which at tree level
contain
the additional parameters $A_\lambda$ and $A_k$ of the
soft symmetry breaking potential of the NMSSM
\begin{eqnarray}
\label{soft}
V_{\mbox{\scriptsize soft}} & = &
m_1^2 |H_1|^2 + m_2^2 |H_2|^2+m_3^2 |N|^2
\nonumber \\ & &
+m_Q^2 |\tilde{Q}|^2 + m_U^2 |\tilde{U}|^2 + m_D^2 |
\tilde{D}|^2
\nonumber \\ & &
+m_L^2 |\tilde{L}|^2 + m_E^2 |\tilde{R}|^2
\nonumber \\  & &
-(\lambda A_\lambda  H_1 H_2 N + \mbox{h.c.})
-(\frac{1}{3}kA_k N^3 + \mbox{h.c.})
\nonumber \\  & &
+ (h_U A_U \tilde{Q} \tilde{U} H_2 -
h_D A_D \tilde{Q} \tilde{D} H_1 -
h_E A_E \tilde{L} \tilde{R} H_1
+\mbox{h.c.})
\nonumber \\ & &
+\frac{1}{2}M \lambda^a \lambda^a
+\frac{1}{2}M' \lambda ' \lambda ' \; .
\end{eqnarray}
In eq.~(\ref{soft})
generation indices are suppressed and the notation of the $SU(2)$
doublet and $U(1)$ singlet fields is conventional.

The NMSSM offers an extremely interesting and complex variety of
neutralino and Higgs phenomenology different from the
minimal model \cite{ellwanger,elliott,kim,king}.
In this paper we analyze the production of neutralinos
$e^+e^- \longrightarrow \tilde{\chi}^0_i  \tilde{\chi}^0_j$
($i,j=1,\ldots ,5$) in the NMSSM. We focus on center-of-mass energies
between 100 and 600 GeV, which cover the energy range of LEP2 up to
that of a planned linear collider. In order to determine the supersymmetric
signatures in the NMSSM we then discuss the subsequent neutralino decays
into fermions, Higgs bosons and photons and compute the dominant decay
channels in seven typical scenarios.
For completeness we add an appendix with the neutralino and Higgs mixing
matrices in the NMSSM and the relevant formulae for neutralino
production and decay.

\section{Scenarios}
In this section we describe the seven scenarios A -- G in which production
and decay of neutralinos in the NMSSM are studied. They
clearly differ from the MSSM by giving at least one of the light
neutralinos a significant singlet component. Also we consider
different values for the gaugino mass parameter $M$ and the
singlet vacuum expectation value $x$ in order to cover various
typical regions in the $(M,x)$-plane.
Furthermore for the decays of the NMSSM neutralinos the allowed
mass regions for the light scalar and pseudoscalar Higgs bosons
are of great importance. Since contrary to the MSSM the Higgs sector
of the NMSSM is strongly correlated to the neutralino sector \cite{franke2},
the masses of the Higgs bosons are bounded already by fixing the parameters
of the neutralino mass matrix.

Concerning the mass of the lightest neutralino which we assume to be
the lightest supersymmetric particle (LSP) we choose on the one hand the
scenarios A -- D where the light neutralino has a mass
of 10 GeV well below
the mass bound for a MSSM neutralino, but we consider on the other hand also
the scenarios E -- G with a 50-GeV LSP. In order to demonstrate
the differences between NMSSM and MSSM, in the first case the singlet
component of the LSP $|\! \! <\!\!\chi^0_1 | \psi_N\!\! >\! \! |^2$
is larger than \mbox{90 \%}, while in the second case the second lightest
neutralino has such a large singlet component
$|\! \! <\!\!\chi^0_2 | \psi_N\!\! >\! \! |^2 > 0.9$.
In all scenarios we restrict ourselves to a single value
$\tan\beta=2$.

Since we want to study the fundamental differences of neutralino production
and decay in MSSM and NMSSM over a wide but for the NMSSM typical
range of parameters we do not consider special solutions of the
renormalization group equations of the NMSSM \cite{renor} as implied by
supergravity models. Also we do not explicitly address the dark matter problem
which has been studied in ref.~\cite{dark} assuming the LSP to be the main
component of dark matter.

\begin{figure}
\begin{center}
\begin{picture}(16,17)
\put(-0.5,13){\includegraphics{nfig1a.ps}}
\put(2.7,18.5){\large $M=200$ GeV, $x=1000$ GeV,
$\tan\beta=2$}
\put(-0.3,17.7){$k$}
\put(7.0,12.7){$\lambda$}
\put(2.7,13.5){\tiny 10 GeV}
\put(3.7,13.9){\tiny 30 GeV}
\put(4.7,14.3){\tiny 50 GeV}
\put(8.0,13){\includegraphics{nfig1b.ps}}
\put(8.2,17.7){$k$}
\put(15.5,12.7){$\lambda$}
\put(11.5,14.5){\tiny 90 \% }
\put(12.5,15.3){\tiny 1 \% }
\put(13.5,16.0){\tiny 0.5 \% }
\put(14.5,16.9){\tiny 0.1 \% }
\put(-0.5,6.5){\includegraphics{nfig1c.ps}}
\put(2.7,12.0){\large $M=120$ GeV, $x=1000$ GeV,
$\tan\beta=2$}
\put(-0.3,11.2){$k$}
\put(7.0,6.2){$\lambda$}
\put(2.7,7.0){\tiny 10 GeV}
\put(3.7,7.4){\tiny 30 GeV}
\put(4.7,8.0){\tiny 50 GeV}
\put(8.0,6.5){\includegraphics{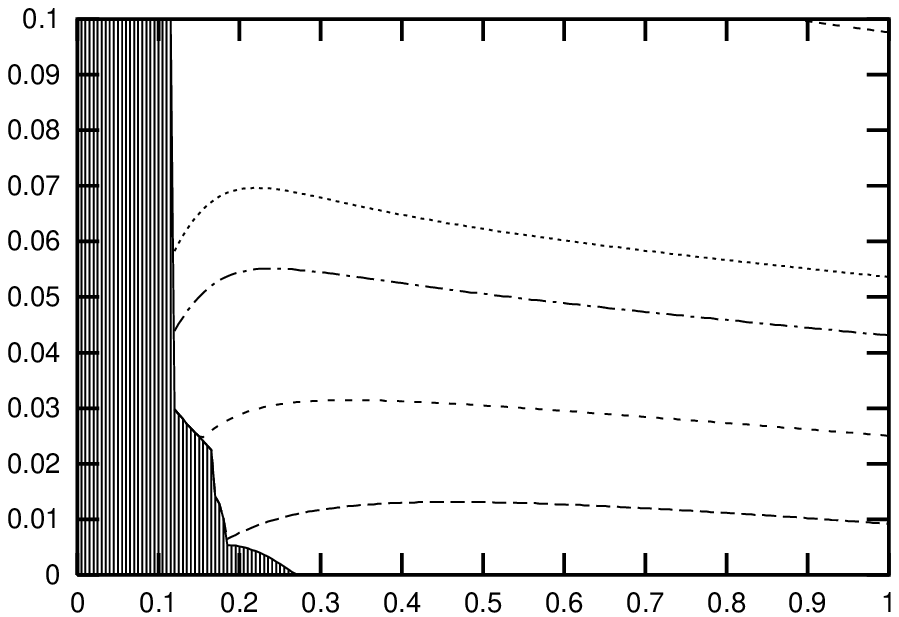}}
\put(8.2,11.2){$k$}
\put(15.5,6.2){$\lambda$}
\put(11.5,7.2){\tiny 90 \% }
\put(12.5,7.8){\tiny 1 \% }
\put(13.5,8.6){\tiny 0.5 \% }
\put(14.5,9.1){\tiny 0.1 \% }
\put(-0.5,0){\includegraphics{nfig1e.ps}}
\put(-0.3,4.7){$k$}
\put(7.0,-0.3){$\lambda$}
\put(3.0,0.6){\tiny 10 GeV}
\put(5.2,4.3){\tiny 30 GeV}
\put(8.0,0){\includegraphics{nfig1f.ps}}
\put(2.7,5.5){\large $M=65$ GeV, $x=1000$ GeV,
$\tan\beta=2$}
\put(8.2,4.7){$k$}
\put(15.5,-0.3){$\lambda$}
\put(11.5,0.9){\tiny 50 \% }
\put(12.5,1.8){\tiny 1 \% }
\put(13.5,2.2){\tiny 0.5 \% }
\put(14.5,4.1){\tiny 0.1 \% }
\end{picture}
\end{center}
\caption{Contour lines of the mass $m_{\tilde{\chi}^0_1}$ and singlet
component $|\!\! <\!\! \psi_N | \chi^0_1 \!\! >\!\! |^2$
of the lightest neutralino in the
$(\lambda ,k)$-plane. The experimentally excluded domain is shaded.}
\label{mlsp}
\end{figure}

First we present in Fig.~\ref{mlsp} the parameter regions in the
($\lambda,k$)-plane of the trilinear couplings in the superpotential
which lead to a
very light or a singlet-like LSP. For three different values of
the gaugino mass parameter $M=65,$ $120,$ and $200$ GeV, the contour lines
for the mass $m_{\tilde{\chi}_1^0}$ of the lightest neutralino and its
singlet component $|\! \! <\!\!\chi^0_1 | \psi_N\!\! >\! \! |^2$
are shown. Since only large singlet vacuum expectation values $x$
allow for singlet components above \mbox{90 \%}, we set $x=1000$ GeV.
The experimentally excluded parameter space (for details see
ref.~\cite{franke1}) is shaded. Note that the $k$ axis ends at 0.1
since only for small couplings $k \lsim 10^{-2}$ the mass of the LSP
in the NMSSM can lie below the experimental bounds of the MSSM.
For most of the ($\lambda,k$)-plane the LSP is heavier than 30 GeV.
Larger masses for the LSP can be obtained with a broad range of
$k$ values, but for large couplings $k \gsim 0.1$ the coupling
$\lambda$ is limited to a narrow interval.

Generally, for smaller
parameters $M$ light neutralinos can exist in a larger region of
the parameter space. The mass range approximately follows
from the asymptotical values at large singlet vacuum expectation
values $x$ \cite{pandita},
$m_{\tilde{\chi}_i^0} \approx -\alpha M,-M,\lambda x,
-\lambda x, -2kx$. Due to $m_{\tilde{\chi}^0_1} \lsim \alpha M$
it is obvious that e.~g.~for $M=65$ GeV and $x=1000$ GeV a LSP with a mass
as large as 50 GeV cannot exist.

Small couplings of the order ${\cal O}(10^{-2})$ are also necessary
to obtain a large singlet component in the mixing of the lightest
neutralino. The requirement of small couplings $k$ for large
singlet components is weakened for larger parameters $M$. While for
$M=200$ GeV a coupling $k \approx 0.3$ still leads to a singlet
component $|\!\! <\!\! \psi_N | \chi^0_1\!\! >\!\! |^2 > 0.9$,
such a large singlet component is impossible with $M=65$ GeV.

The scenarios A -- D are constructed under the prerequisite
that the lightest neutralino has a
mass of about 10 GeV.
For that we fix the couplings $\lambda=0.4$ and $k=0.001$ and
keep in the scenarios A -- C the parameters of Fig.~\ref{mlsp}:
$\tan\beta=2$, $x=1000$ GeV, $M=65$, $120,$ $200$ GeV. With
scenario D we want to study the consequences of a negative
gaugino mass parameter $M=-120$ GeV. The masses and mixings of the
neutralinos in these scenarios are given in Tables \ref{szetabab} and
\ref{szetabcd}. They mainly differ by the sign of the neutralino
mass eigenvalues, the masses of the next lightest
neutralinos and the chargino masses.

With the fixed values of $\lambda$, $k$, $x$ and $\tan\beta$
and scanning over the parameters $A_\lambda$ and $A_k$ of the
Higgs sector also
the mass ranges for the scalar and pseudoscalar Higgs bosons and
their mixings are identical in the scenarios A -- D and shown in
Table \ref{szetabhiggsa}. Here also the experimental mass bounds as
described in ref.~\cite{franke2} are included.
Therefore, in our discussion of the neutralino decays
in the scenarios A -- D we may focus on the consequences of the different
neutralino mass eigenvalues and mixings and can disregard
variations of the Higgs masses. Because of this advantage we tolerate
the slightly different masses of the LSP in these scenarios which,
however, should not significantly alter the numerical results.
The charged Higgs bosons are as heavy as 1 TeV in these scenarios and
are therefore not relevant for the neutralino decays.

In the scenarios E -- G (Tables \ref{szetabef} and \ref{szetabg})
the lightest neutralino
has a mass of 50 GeV. As it can be seen in Fig.~\ref{mlsp}, this mass
value requires larger couplings $k$ than in the scenarios A -- D.
Then also the singlet component of the LSP decreases, so that now
the second lightest neutralino is mainly a singlet.
Also in these scenarios with a heavier LSP we want to analyze the effect
of the sign of the gaugino mass parameter $M$. In scenario E we first
choose a positive value. But
since negative gaugino mass parameters M allow for larger singlet
components of the second lightest neutralino,
the scenarios F and G are constructed with $M=-95$ GeV and $M=-90$ GeV,
respectively,
so that the differences to the MSSM are emphasized.
While in scenarios E and F we retain the large singlet vacuum expectation value
$x=1000$ GeV with $\lambda=0.4$, $k=0.035$, $\tan\beta=2$, we consider
in scenario G a smaller singlet value $x=300$ GeV and $\lambda=0.4$, $k=0.1$,
$\tan\beta=2$. This smaller $x$ value leads to an increase of the doublet
higgsino components of the light neutralinos and diminishes their
zino components.

Also shown in Tables \ref{szetabef} and \ref{szetabg}
is the allowed mass range for the
scalar and pseudoscalar Higgs bosons scanning over the parameters
$A_\lambda$ and $A_k$ of the Higgs sector. Since in both scenarios
the mass differences between the light neutralinos is chosen to be
rather small in order enhance production cross sections even at
LEP2 energies, the decay $\tilde{\chi}_2^0 \longrightarrow
\tilde{\chi}_1^0 +${\sl Higgs} is kinematically not possible in
scenarios E and F due to the large lower Higgs mass bound.
We will discuss the neutralino decays in the NMSSM in detail later.
Now we first focus on the neutralino production in the above described
scenarios.
\section{Production of NMSSM neutralinos}
In this section we discuss the production of neutralinos at
electron-positron colliders. Here we compute the cross sections for
neutralino production in the seven scenarios and determine the
center-of-mass energy necessary for the identification of a NMSSM
neutralino. An analysis of the possible signatures from
the decays follows in the next section.

Neutralino production $e^+e^- \longrightarrow \tilde{\chi}_i^0
\tilde{\chi}_j^0$ ($i,j=1,\ldots,5$) proceeds via $Z$-exchange in the
$s$ channel and exchange of a selectron in the $t$ and $u$ channels.
The corresponding Feynman graphs and
the relevant $eeZ$, $\tilde{\chi}_i^0\tilde{\chi}_j^0Z$ and
$e\tilde{e}\tilde{\chi}_i^0$ vertex factors
are shown in Fig.~\ref{graph}.
The notation is explained in the appendix where also the analytical
formulae for
the cross section are collected.

Since the singlet superfield has hypercharge 0, the singlet component of
the neutralinos does not couple to (s)fermions and gauge bosons, so that
the analytical expressions for neutralino production in the NMSSM are
identical to those in the minimal model. The nevertheless often drastic
differences merely arise by the mixings of the neutralinos.
Before we analyze neutralino production in the previously presented
scenarios we first want to make clear characteristic differences
between MSSM and NMSSM with two simple examples.
Let the MSSM be realized in the nature and imagine a MSSM
scenario where only the LSP can be produced at a collider,
e.~g.~LEP2, since the other neutralinos are too heavy. Then
it would be impossible to detect a neutralino
due to R-parity conservation. If, however, nature is described by
the NMSSM, there could exist
an additional light neutralino as LSP, so that the a neutralino
with a similar mass but invisible in the MSSM
could be identified by its decay into the LSP.

While in this example a NMSSM scenario could be verified more easily than
the corresponding MSSM scenario, even the contrary could be possible: In the
NMSSM the production of a singlet-like second lightest neutralinos
could be heavily suppressed leading to two practically invisible
neutralinos.

\begin{figure}
\begin{center}
\begin{picture}(16,18.5)
\put(0,14){\includegraphics{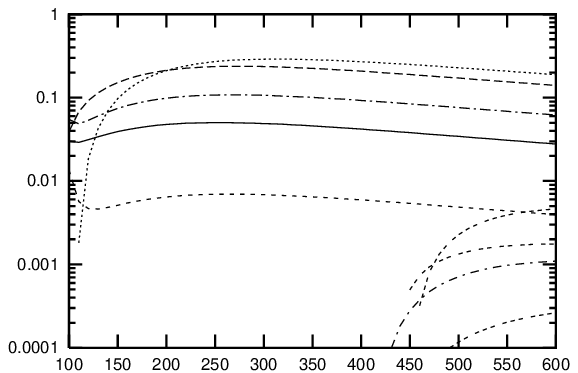}}
\put(5.0,13.5){$\sqrt{s}/$GeV}
\put(4.0,14.7){Scenario A}
\put(-0.0,20.0){$\sigma/$pb}
\put(0,7){\includegraphics{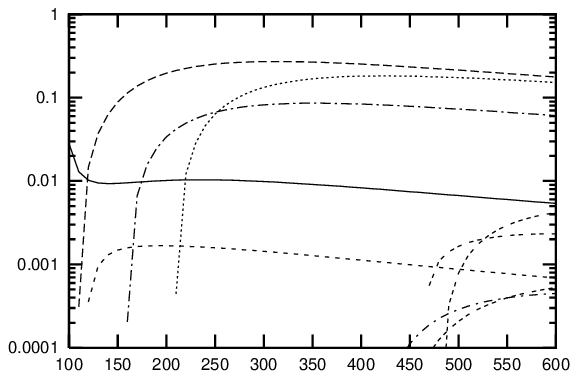}}
\put(5.0,6.5){$\sqrt{s}/$GeV}
\put(-0.0,13.0){$\sigma/$pb}
\put(4.0,7.7){Scenario B}
\put(0,0){\includegraphics{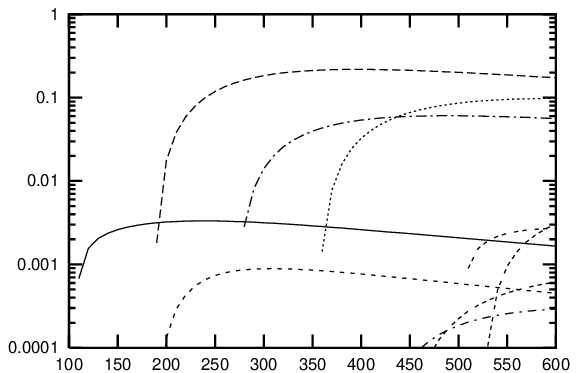}}
\put(5.0,-0.5){$\sqrt{s}/$GeV}
\put(-0.0,6.0){$\sigma/$pb}
\put(4.0,0.7){Scenario C}
\put(9,6.5){\includegraphics{beschr.ps}}
\put(12.5,13.3){$e^+e^- \rightarrow \tilde{\chi}^0_1 \tilde{\chi}^0_2$}
\put(12.5,12.5){$e^+e^- \rightarrow \tilde{\chi}^0_1 \tilde{\chi}^0_3$}
\put(12.5,11.7){$e^+e^- \rightarrow \tilde{\chi}^0_2 \tilde{\chi}^0_2$}
\put(12.5,11.0){$e^+e^- \rightarrow \tilde{\chi}^0_2 \tilde{\chi}^0_3$}
\put(12.5,10.2){$e^+e^- \rightarrow \tilde{\chi}^0_3 \tilde{\chi}^0_3$}
\put(12.5,9.4){$e^+e^- \rightarrow \tilde{\chi}^0_1 \tilde{\chi}^0_4$}
\put(12.5,8.7){$e^+e^- \rightarrow \tilde{\chi}^0_2 \tilde{\chi}^0_4$}
\put(12.5,7.9){$e^+e^- \rightarrow \tilde{\chi}^0_1 \tilde{\chi}^0_5$}
\put(12.5,7.1){$e^+e^- \rightarrow \tilde{\chi}^0_2 \tilde{\chi}^0_5$}
\end{picture}
\end{center}
\caption{Cross sections for
$e^+e^- \longrightarrow \tilde{\chi}^0_i \tilde{\chi}^0_j$
in scenarios A -- C.}
\label{prodfiga}
\end{figure}

\begin{figure}
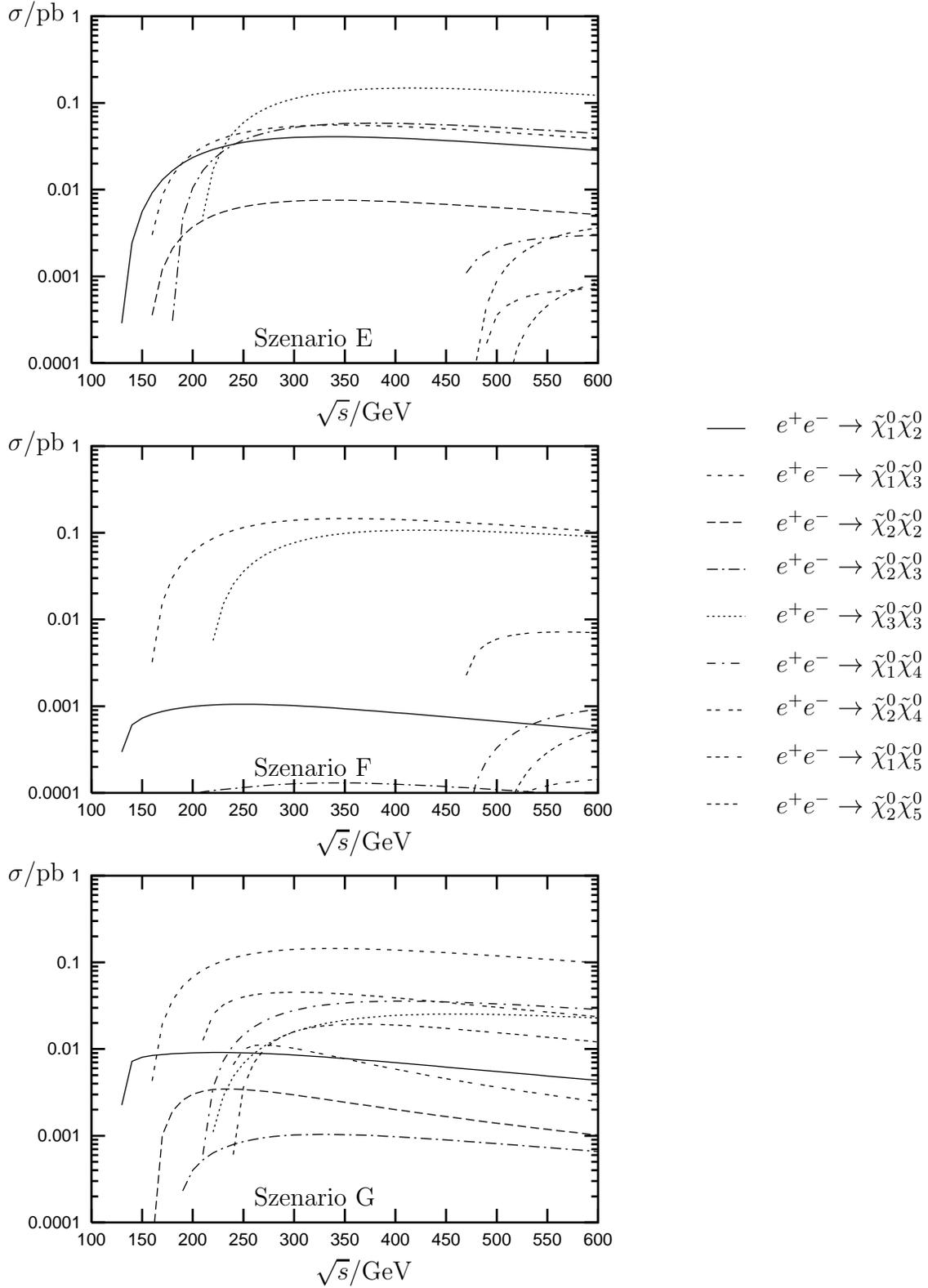

\begin{center}
\begin{picture}(16,18.5)
\put(0,14){\includegraphics{nfig5a.ps}}
\put(5.0,13.5){$\sqrt{s}/$GeV}
\put(4.0,14.7){Szenario E}
\put(-0.0,20.0){$\sigma/$pb}
\put(0,7){\includegraphics{nfig5b.ps}}
\put(5.0,6.5){$\sqrt{s}/$GeV}
\put(-0.0,13.0){$\sigma/$pb}
\put(4.0,7.7){Szenario F}
\put(0,0){\includegraphics{nfig5c.ps}}
\put(5.0,-0.5){$\sqrt{s}/$GeV}
\put(-0.0,6.0){$\sigma/$pb}
\put(4.0,0.7){Szenario G}
\put(9,6.5){\includegraphics{beschr.ps}}
\put(12.5,13.3){$e^+e^- \rightarrow \tilde{\chi}^0_1 \tilde{\chi}^0_2$}
\put(12.5,12.5){$e^+e^- \rightarrow \tilde{\chi}^0_1 \tilde{\chi}^0_3$}
\put(12.5,11.7){$e^+e^- \rightarrow \tilde{\chi}^0_2 \tilde{\chi}^0_2$}
\put(12.5,11.0){$e^+e^- \rightarrow \tilde{\chi}^0_2 \tilde{\chi}^0_3$}
\put(12.5,10.2){$e^+e^- \rightarrow \tilde{\chi}^0_3 \tilde{\chi}^0_3$}
\put(12.5,9.4){$e^+e^- \rightarrow \tilde{\chi}^0_1 \tilde{\chi}^0_4$}
\put(12.5,8.7){$e^+e^- \rightarrow \tilde{\chi}^0_2 \tilde{\chi}^0_4$}
\put(12.5,7.9){$e^+e^- \rightarrow \tilde{\chi}^0_1 \tilde{\chi}^0_5$}
\put(12.5,7.1){$e^+e^- \rightarrow \tilde{\chi}^0_2 \tilde{\chi}^0_5$}
\end{picture}
\end{center}
\caption{Cross sections for
$e^+e^- \longrightarrow \tilde{\chi}^0_i \tilde{\chi}^0_j$
in the scenarios E -- G.}
\label{prodfige}
\end{figure}

After this qualitative consideration we now come to the numerical results.
Figs.~\ref{prodfiga} -- \ref{prodfige} show the cross sections for
neutralino production
$e^+e^- \longrightarrow \tilde{\chi}_i^0
\tilde{\chi}_j^0$ ($i,j=1,\ldots,5$) as a function of the center-of-mass
energy in the range $100$ GeV $\leq \sqrt{s} \leq 600$ GeV covering
the LEP2
energy $\sqrt{s}\approx 190$ GeV and the energy
linear collider $\sqrt{s}=500$ GeV of a future linear collider.
Especially we want to determine the
energy range where the cross sections in typical NMSSM scenarios reach
magnitudes large enough in order to have reasonable chances for a
detection.

All cross sections are computed with masses $m_{\tilde{e}_{L,R}}=200$ GeV
for the left-handed and right-handed selectron. This value lies well above the
current mass bounds but is not so large that the production rates
are heavily suppressed. We have neglected the mass splitting between
left and right-handed selectrons as it appears within a unified theory
\cite{polchinski},
which, however, does not significantly alter the numerical results.
The dependence of the cross sections on the selectron mass has already been
studied in ref.~\cite{kim}. For $m_{\tilde{e}_{L,R}}=100$ GeV they increase
by a factor of 2 -- 5, for $m_{\tilde{e}_{L,R}}=1000$ they decrease by one
order of magnitude.

For the considered energy range above the $Z$-peak the cross sections are
dominated by selectron exchange. Therefore the dependence on the
center-of-mass energy is similar for all scenarios: After the kinematical
threshold there begins a steep increase to the maximum which
is followed by a flat decrease.
The experimental identification of a neutralino may be facilitated
in typical NMSSM scenarios with a light singlet-like LSP, where
the next lightest neutralino is mainly composed of photinos and zinos.
Since the selectron couple only to
the photino/zino components of the
neutralinos, this characteristic
mixing type in our scenarios favors the production of the light neutralinos.

In scenario A the production of the
three lightest neutralino is kinematically possible even for moderate
collider energies and proceeds with cross sections above 10 fb in the
channels $e^+e^- \longrightarrow \tilde{\chi}^0_1 \tilde{\chi}^0_2,
\tilde{\chi}^0_2 \tilde{\chi}^0_2,
\tilde{\chi}^0_2 \tilde{\chi}^0_3,
\tilde{\chi}^0_3 \tilde{\chi}^0_3$. The
maximum lies around energies of 250 -- 300 GeV but is approached already
at the expected LEP2 energy of 190 GeV. For this energy
the cross sections for pair production of the second and third lightest
neutralino both reach 200 fb.
In scenario A, the light chargino
has a mass significantly below 100 GeV so that there are good chances to
detect it at LEP2. Then pair production of $\tilde{\chi}^0_2$
followed by the decay into the LSP offers
the possibility to discriminate between NMSSM and MSSM.

In scenario B with the larger gaugino mass parameter $M$, only the
cross sections for
$\tilde{\chi}^0_2 \tilde{\chi}^0_2$ and $\tilde{\chi}^0_2 \tilde{\chi}^0_3$
production are above 10 fb at $\sqrt{s}=190$ GeV.
For larger energies
$\sqrt{s} \gsim 250$ GeV also pair production of $\tilde{\chi}^0_3$
reach values of about 100 fb.
With increasing parameter $M$ in scenario C finally only $\tilde{\chi}^0_1
\tilde{\chi}^0_2$ production is kinematically possible at LEP2 energies
but is heavily suppressed due to the strong singlet component of the LSP.
This scenario makes clear, that a light NMSSM neutralino with mass
of 10 GeV or below does not
necessarily need to be detected at LEP2, while a
500-GeV linear collider definitely seems capable for verification or
exclusion of a very light NMSSM neutralino provided that $M$ is bounded by the
fine-tuning or naturalness constraint
\begin{equation}
\label{hierarchie}
-400 \; \mbox{GeV} \le M \le 400 \; \mbox{GeV}.
\end{equation}
This estimation arises by assuming a mass bound of about 1 TeV for
all supersymmetric particles, especially the gluino, with
\begin{equation}
|M|  =  \frac{\alpha_2}{\alpha_3}m_{\tilde{g}} \simeq 0.3
m_{\tilde{g}} ,
\label{mgluino}
\end{equation}
where the $\alpha_i$ are the gauge couplings of the symmetry groups.

Scenario D differs from scenario B mainly by the sign of the
gaugino mass parameter and therefore by the relative signs of the mass
eigenvalues of the light neutralinos.
Further the singlet component of the LSP is slightly higher for
negative $M$. Both lead to a decrease of the cross sections for
production of the LSP together with another neutralino and an increase of the
$\tilde{\chi}^0_2$ pair production due to the larger photino/zino
component of the second lightest neutralino in scenario D compared to
scenario B.

Since in the scenarios E -- G the LSP already has a mass of 50 GeV,
the production rates are rather low at LEP2 energies.
While in scenario E the production of $\tilde{\chi}^0_2$ and
$\tilde{\chi}^0_3$ together with the LSP proceeds at $\sqrt{s}=190$ GeV
with cross section
above 10 fb, in scenarios F and G
the cross sections for the production of the LSP
with the third neutralino are around 100 fb. Here $\tilde{\chi}^0_1
\tilde{\chi}^0_2$ production is favored in scenario E compared to
the scenarios F and G because the singlet component of the second neutralino
is reduced for positive parameters $M$, and the
$\tilde{\chi}^0_1 \tilde{\chi}^0_3$ production is suppressed due to the
larger singlet component of the LSP. Even at higher energies the
channels for neutralino detection are significantly reduced. Again only
the third lightest neutralino is produced with rates of 100 fb
in these scenarios, but in scenarios E and G also the second neutralino
reaches cross section above 10 fb at high energies in the range of a
linear collider.

Finally we consider in scenario G the case of smaller $x$ values.
While the neutralino production at $\sqrt{s}=190$ GeV hardly differs
from scenario E, this situation changes already at slightly higher
energies $\sqrt{s} \gsim 200$ GeV: Now a variety of production
channels opens, even the heavy neutralinos are produced with cross sections
of about 30 -- 50 fb. In such a scenario the interpretation of the
experimental data and the verification of a concrete NMSSM scenario get
rather complicated due to cascade decays of the heavy neutralinos.

The scenarios E -- G show that the production of the singlet-like
second lightest neutralino is heavily suppressed. Nevertheless the cross
section at a future linear collider may be sufficient for an
identification by its subsequent decay into the LSP. Another
possibility for a distinction of NMSSM and MSSM scenarios would be
the decays of the
heavier neutralinos into the singlet-like second neutralino.
This is the subject of the next section.
\section{Decay of NMSSM neutralinos}
While the lightest neutralino as the LSP is stable and invisible
in the MSSM as well
as in the NMSSM, the other neutralino decay and can be detected in principle
by their decay products. Since at the end of each decay chain the undetectable
LSP is produced, missing energy is one essential supersymmetric signature.
The heavy neutralinos can decay directly into the LSP, but also via
numerous cascade decays. We restrict ourselves in this paper on the
decays of the light neutralinos which are produced with a sufficiently
large cross section. The following decays could represent the begin of a
decay chain:
\begin{enumerate}
\item the decay of a heavy neutralino into a lighter one and two fermions
$\tilde{\chi}^0_i \longrightarrow \tilde{\chi}^0_j e^+e^-$,
$\tilde{\chi}^0_j \nu \bar{\nu}$, $\tilde{\chi}^0_j q \bar{q}$ ($i>j$).

Since we assume these fermions to be massless, these decays are always
kinematically allowed. They proceed via $Z$ and sfermion exchange, their
respective Feynman graphs are shown in Fig.~\ref{zerfey} and the decay
widths are given in the appendix. If one of the neutralinos has a large
singlet component, the decay width for these three body decays gets
rather small compared to the MSSM. We will emphasize this fact in the
further discussion of the decays.
\item the decay of a heavy neutralino into a chargino and two fermions
$\tilde{\chi}^0_i \longrightarrow \tilde{\chi}^\pm_j e^\pm \nu$,
$\tilde{\chi}^\pm_j qq'$.

In our scenarios the third neutralino is a few GeV heavier than the light
chargino, so that this decay becomes kinematically possible. Due to the
small mass difference it is, however, strongly suppressed compared to
the three body decay into a light neutralino. Therefore we neglect this
decay channel for the rest of this paper.
\item the decay of a heavy neutralino into a light neutralino and a photon
$\tilde{\chi}^0_i \longrightarrow \tilde{\chi}^0_j \gamma$
($i>j$).

This decay is always kinematically allowed. Since it proceeds in
lowest order via a loop with $W$ bosons, charginos and charged Higgs
bosons or sfermions, it is suppressed compared to the
tree level decays. The complete set of Feynman graphs can be found
in ref.~\cite{wyler}, we show in Fig.~\ref{zerfey} only those with
neutralino-chargino-Higgs couplings which differ from the MSSM.

Generally, the decay width of this loop decay is rather small. Therefore,
in the MSSM it is relevant only if one of the neutralinos is nearly a pure
photino and the other mainly a higgsino, so that the three body decay is also
suppressed. In the NMSSM, however, both decay widths could be of the same
order of magnitude. This is the case if one of the neutralinos has a
large singlet component which explicitly affects the neutralino-chargino-Higgs
coupling of the loop decay
into a photon but just reduces the three body decay rates.
\item the decay of a heavy neutralino into a light neutralino and a
scalar or pseudoscalar neutral Higgs boson
$\tilde{\chi}^0_i \longrightarrow \tilde{\chi}^0_j
S_a,P_b$.

This decay proceeds at tree level via the Feynman graphs shown in
Fig.~\ref{zerfey} with the corresponding analytical formulae for
the decay width given in the appendix. The Higgs-neutralino-neutralino
coupling is significantly influenced by the singlet components of the
neutralinos and the Higgs boson. In our scenarios with a light singlet-like
neutralino there often exist also light neutral Higgs bosons, so that
this decay becomes kinematically possible and dominant, while in the
MSSM the actual Higgs mass bounds forbid such a light Higgs particle.
\item the decay of a heavy neutralino into a light neutralino or a chargino
and a gauge boson, into a fermion and sfermion or into a chargino
together with a charged Higgs boson.

Since in our scenarios the mass difference between the third neutralino
and the LSP is maximal about 45 GeV, we do not consider these decays in
the following.
\end{enumerate}

If the decay $\tilde{\chi}^0_i \longrightarrow \tilde{\chi}^0_j
S_1/P_1$ dominates, the signatures of neutralino production crucially
depend on the decay mechanism of the produced Higgs boson. Possible
decay products are according to the masses and mixings two heavy quarks or
leptons via $S_1/P_1 \longrightarrow b\bar{b},\tau\bar{\tau}$ or two
neutralinos $S_1/P_1 \longrightarrow \tilde{\chi}^0_1 \tilde{\chi}^0_1$.
In the first case a clear signature may arise if the standard model
background can be suppressed, in the second case the neutralino decay
into the LSP plus a Higgs decay cannot be
detected.
Further, the channel $S_1 \longrightarrow P_1 P_1$ could be the
starting point for a cascade decay of the lightest Higgs scalar.

\begin{figure}
\begin{center}
\begin{picture}(16,19)
\put(0.5,14){\includegraphics{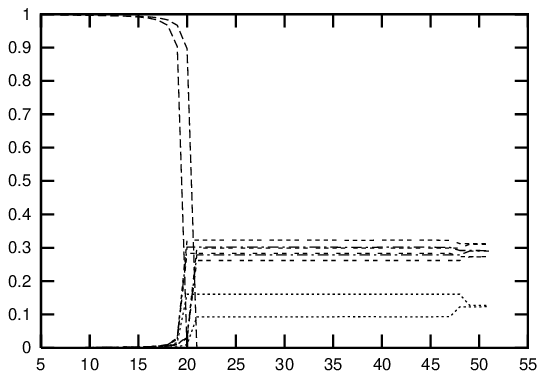}}
\put(5.5,13.5){$m_{P_1}/$GeV}
\put(0.5,7){\includegraphics{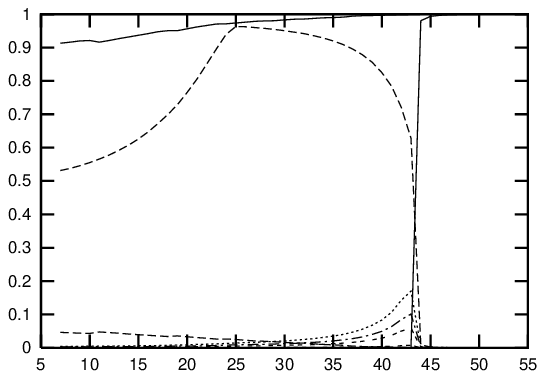}}
\put(5.5,6.5){$m_{P_1}/$GeV}
\put(0.5,0){\includegraphics{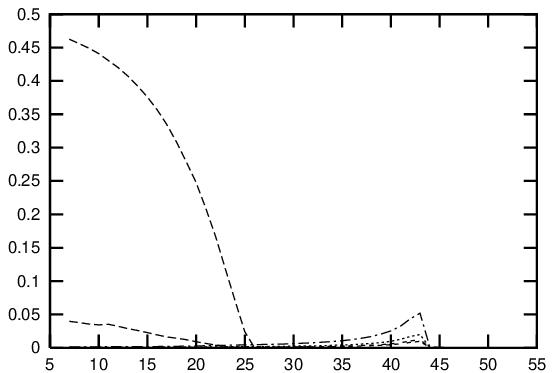}}
\put(5.5,-0.5){$m_{P_1}/$GeV}
\put(9.0,12.0){\includegraphics{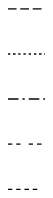}}
\put(12.5,18.8){$\tilde{\chi}^0_2 \rightarrow \tilde{\chi}^0_1 P_1$}
\put(12.5,18.0){$\tilde{\chi}^0_2 \rightarrow \tilde{\chi}^0_1 \gamma$}
\put(12.5,17.2){$\tilde{\chi}^0_2 \rightarrow \tilde{\chi}^0_1 q\bar{q}$}
\put(12.5,16.5){$\tilde{\chi}^0_2 \rightarrow \tilde{\chi}^0_1 \nu\bar{\nu}$}
\put(12.5,15.8){$\tilde{\chi}^0_2 \rightarrow \tilde{\chi}^0_1 e^+e^-$}
\put(9,5.5){\includegraphics{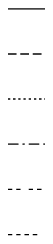}}
\put(12.5,12.3){$\tilde{\chi}^0_3 \rightarrow \tilde{\chi}^0_1 S_1$}
\put(12.5,11.5){$\tilde{\chi}^0_3 \rightarrow \tilde{\chi}^0_1 P_1$}
\put(12.5,10.7){$\tilde{\chi}^0_3 \rightarrow \tilde{\chi}^0_1 \gamma$}
\put(12.5,10.0){$\tilde{\chi}^0_3 \rightarrow \tilde{\chi}^0_1 q\bar{q}$}
\put(12.5,9.3){$\tilde{\chi}^0_3 \rightarrow \tilde{\chi}^0_1 \nu\bar{\nu}$}
\put(12.5,8.5){$\tilde{\chi}^0_3 \rightarrow \tilde{\chi}^0_1 e^+e^-$}
\put(9,-2.0){\includegraphics{beschr24579.ps}}
\put(12.5,4.8){$\tilde{\chi}^0_3 \rightarrow \tilde{\chi}^0_2 P_1$}
\put(12.5,4.0){$\tilde{\chi}^0_3 \rightarrow \tilde{\chi}^0_2 \gamma$}
\put(12.5,3.2){$\tilde{\chi}^0_3 \rightarrow \tilde{\chi}^0_2 q\bar{q}$}
\put(12.5,2.5){$\tilde{\chi}^0_3 \rightarrow \tilde{\chi}^0_2 \nu\bar{\nu}$}
\put(12.5,1.8){$\tilde{\chi}^0_3 \rightarrow \tilde{\chi}^0_2 e^+e^-$}
\end{picture}
\end{center}
\caption{Minima und maxima of the branching ratios of the neutralino
decays in scenario A as a function of the mass of the light pseudoscalar
Higgs boson.}
\label{zerfiga}
\end{figure}

We now discuss the neutralino decays in our scenarios.
First we consider in scenario A the decays of the second and third
neutralino that can be produced at LEP2 or a linear collider.
The branching ratios for these decays are depicted in Fig.~\ref{zerfiga}
as a function of the mass of the light pseudoscalar Higgs boson.
Since the lightest scalar Higgs boson has a
mass of at least 37 GeV in the scenarios A -- D,
it can be produced only in the direct decay
of the third neutralino into the LSP, then with branching rations between
0 and 1 according to the choice of the parameters $A_\lambda$ and $A_k$.

\begin{figure}
\begin{center}
\begin{picture}(16,6)
\put(0.5,0){\includegraphics{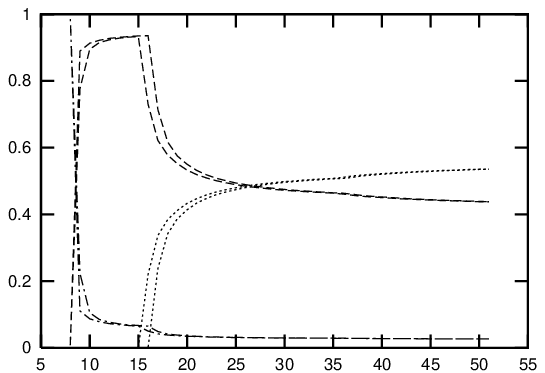}}
\put(5.5,-0.5){$m_{P_1}/$GeV}
\put(9,-2.8){\includegraphics{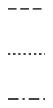}}
\put(12.5,4.0){$P_1 \rightarrow b \bar{b}$}
\put(12.5,3.2){$P_1 \rightarrow \tilde{\chi}^0_1 \tilde{\chi}^0_1$}
\put(12.5,2.5){$P_1 \rightarrow  \tau\bar{\tau}$}
\end{picture}
\end{center}
\caption{Minima und Maxima of the branching ratios of the decays of the light
pseudoscalar Higgs boson in scenario A.}
\label{higgszerap}
\end{figure}

We show in Fig.~\ref{zerfiga} the maximal and minimal branching ratios
scanning over all experimentally allowed values of $A_\lambda$ and $A_k$.
If the mass of the light pseudoscalar Higgs boson is smaller than the mass
difference between the two lightest neutralinos, that is about 20 GeV
in scenario A, the second lightest neutralino decays nearly completely
into the LSP and $P_1$. Then the signatures of $\tilde{\chi}^0_2$
depend on the dominant decay channels of the light pseudoscalar Higgs
shown in Fig.~\ref{higgszerap}. For pseudoscalar masses
below 9 GeV the decay into a tau pair dominates. As soon as the decay into
b quarks is kinematically possible, it contributes with at least
\mbox{50 \%}. For $m_{P_1}>16$ GeV also the invisible decay into two LSP
takes place with a branching ratio up to 0.5.

If the pseudoscalar Higgs boson is heavier than 20 GeV, for the second
neutralino the three body decays into two fermions and the LSP
and the loop decay into one photon plus the LSP dominate.
Their branching ratios are shown in Fig.~\ref{zerfiga}  for one
generation. Here the branching ratios of all decay channels are approximately
of the same order, about 0.3 for each three body decay and
up to 0.15 for the loop decay.

\begin{figure}
\begin{center}
\begin{picture}(16,6)
\put(0.5,0){\includegraphics{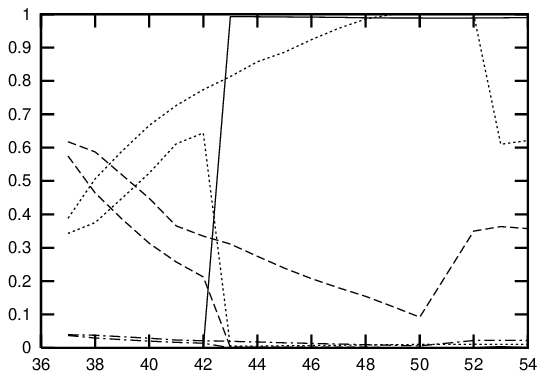}}
\put(5.5,-0.5){$m_{S_1}/$GeV}
\put(9,-2){\includegraphics{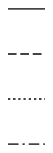}}
\put(12.5,4.8){$S_1 \rightarrow P_1 P_1$}
\put(12.5,4.0){$S_1 \rightarrow b \bar{b}$}
\put(12.5,3.2){$S_1 \rightarrow \tilde{\chi}^0_1 \tilde{\chi}^0_1$}
\put(12.5,2.5){$S_1 \rightarrow \tau\bar{\tau}$}
\end{picture}
\end{center}
\caption{Minima und Maxima of the branching ratios of the decays of the
lightest scalar Higgs boson in scenario A.}
\label{higgszeras}
\end{figure}

Since at least one neutral Higgs particle is lighter than the
mass difference between the third neutralino and the LSP, the
decay of $\tilde{\chi}_3^0$ into a Higgs boson and the LSP
always dominates. Only if the masses of the
scalar and pseudoscalar Higgs boson
are near their upper bounds the other decay modes may
contribute with a maximum of \mbox{60 \%}. The decay channels for the scalar
Higgs boson in scenario A are
$S_1 \longrightarrow b\bar{b},\tilde{\chi}^0_1 \tilde{\chi}^0_1,P_1P_1$
with branching ratios shown in Fig.~\ref{higgszeras}.
The nearly complete decay into two pseudoscalars
is possible, while for other Higgs parameters the invisible decay into two
LSP may clearly dominate.

\begin{figure}
\begin{center}
\begin{picture}(16,19)
\put(0.5,14){\includegraphics{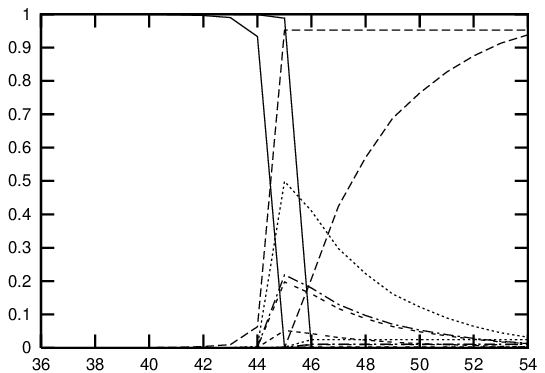}}
\put(5.5,13.5){$m_{S_1}/$GeV}
\put(12.5,19.5){Scenario B}
\put(9,11.5){\includegraphics{beschr124579.ps}}
\put(12.5,18.3){$\tilde{\chi}^0_2 \rightarrow \tilde{\chi}^0_1 S_1$}
\put(12.5,17.5){$\tilde{\chi}^0_2 \rightarrow \tilde{\chi}^0_1 P_1$}
\put(12.5,16.7){$\tilde{\chi}^0_2 \rightarrow \tilde{\chi}^0_1 \gamma$}
\put(12.5,16.0){$\tilde{\chi}^0_2 \rightarrow \tilde{\chi}^0_1 q\bar{q}$}
\put(12.5,15.3){$\tilde{\chi}^0_2 \rightarrow \tilde{\chi}^0_1 \nu\bar{\nu}$}
\put(12.5,14.5){$\tilde{\chi}^0_2 \rightarrow \tilde{\chi}^0_1 e^+e^-$}
\put(0.5,7){\includegraphics{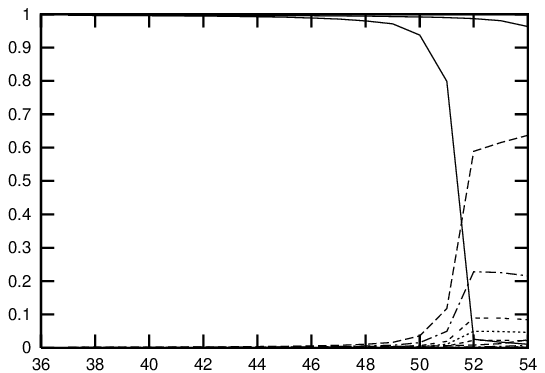}}
\put(5.5,6.5){$m_{S_1}/$GeV}
\put(12.5,12.5){Scenario C}
\put(9,4.5){\includegraphics{beschr124579.ps}}
\put(12.5,11.3){$\tilde{\chi}^0_2 \rightarrow \tilde{\chi}^0_1 S_1$}
\put(12.5,10.5){$\tilde{\chi}^0_2 \rightarrow \tilde{\chi}^0_1 P_1$}
\put(12.5,9.7){$\tilde{\chi}^0_2 \rightarrow \tilde{\chi}^0_1 \gamma$}
\put(12.5,9.0){$\tilde{\chi}^0_2 \rightarrow \tilde{\chi}^0_1 q\bar{q}$}
\put(12.5,8.2){$\tilde{\chi}^0_2 \rightarrow \tilde{\chi}^0_1 \nu\bar{\nu}$}
\put(12.5,7.4){$\tilde{\chi}^0_2 \rightarrow \tilde{\chi}^0_1 e^+e^-$}
\put(0.5,0){\includegraphics{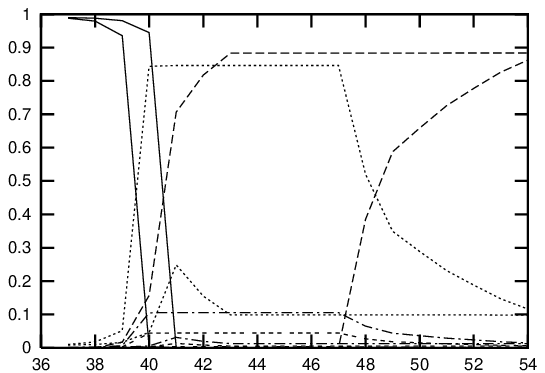}}
\put(5.5,-0.5){$m_{S_1}/$GeV}
\put(12.5,5.5){Scenario D}
\put(9,-2.5){\includegraphics{beschr124579.ps}}
\put(12.5,4.3){$\tilde{\chi}^0_2 \rightarrow \tilde{\chi}^0_1 S_1$}
\put(12.5,3.5){$\tilde{\chi}^0_2 \rightarrow \tilde{\chi}^0_1 P_1$}
\put(12.5,2.7){$\tilde{\chi}^0_2 \rightarrow \tilde{\chi}^0_1 \gamma$}
\put(12.5,2.0){$\tilde{\chi}^0_2 \rightarrow \tilde{\chi}^0_1 q\bar{q}$}
\put(12.5,1.3){$\tilde{\chi}^0_2 \rightarrow \tilde{\chi}^0_1 \nu\bar{\nu}$}
\put(12.5,0.5){$\tilde{\chi}^0_2 \rightarrow \tilde{\chi}^0_1 e^+e^-$}
\end{picture}
\end{center}
\caption{Minima und maxima of the branching ratios of the neutralino
decays in scenarios B -- D as a function of the mass of the lightest
scalar Higgs boson.}
\label{zerfigb}
\end{figure}

In scenario B the mass difference between the two lightest neutralinos
becomes large enough to allow the decay of $\tilde{\chi}^0_2$
into a scalar Higgs boson. Therefore we show in Fig.~\ref{zerfigb}
the branching ratios as a function of the mass of the lightest Higgs
scalar. Again one sees the dominance of the neutralino decays into
Higgs bosons. Only in a small Higgs mass range the other decays
play some important role and reach branching ratios up to
\mbox{50 \%}. Since the singlet component of the lightest neutralino
increases from scenario A to scenario D, also the loop decay
of the second neutralino into the LSP and a photon becomes more
important.

Due to the large mass difference between the third lightest neutralino and
the LSP in the scenarios B -- D, the dominant $\tilde{\chi}^0_3$ decay
mechanism always is the Higgs channel. Here the branching ratios are
very similar to those of scenario A, with the possible subsequent Higgs decays
into a pair of b quarks or two neutralinos according to the
parameters $A_\lambda$ and $A_k$ of the Higgs sector.
Therefore we discuss the $\tilde{\chi}^0_3$ decays not separately in
every scenario B -- D but refer to Figs.~\ref{zerfiga} -- \ref{higgszeras}.

The scenarios E -- G where the LSP has a mass of 50 GeV differ by the
sign of the gaugino mass parameter $M$ and the value
of the singlet vacuum expectation value $x$. In these scenarios we
do not take into consideration neutralino decays into Higgs bosons.
As before these decays dominate if they are kinematically possible, then
the produced Higgs boson decays dominantly into a b quark pair since
the decay into two LSP is not possible due to their large mass.
Instead we give in Table \ref{tabefg} numerical values for the
decay widths and branching ratios for such parameters $A_\lambda$ and
$A_k$ which do not allow neutralino decays into Higgs bosons.
Concerning the decays of the singlet-like second lightest neutralino
we now clearly see the differences between the scenarios.
In scenario E with positive $M$ and large $x$ the leptonic three body
decay dominates, while in scenario F with negative $M$ and large $x$ the
loop decay into photons is dominant and in scenario G with negative $M$
and smaller singlet vacuum expectation value $x$ the hadronic three body
decay is the most important. Due to the large singlet component of the
second neutralino, the decay widths are always very small.

So the MSSM and the NMSSM can be clearly distinguished when a singlet-like
visible neutralino is produced with cross sections large enough for detection.
As discussed in the previous section, this is realized in
scenarios E and G at a 500-GeV linear collider.
The other possibility to verify a NMSSM scenario by the decays of the
heavier neutralinos into the singlet-like second neutralino, however,
seems to be locked. Table \ref{tabefg} shows that similar to the MSSM
the direct decay of $\tilde{\chi}^0_3$ into the LSP dominates by far.
The branching ratios for the significant NMSSM decay into the second
neutralino can be neglected. This decay is strongly suppressed
even in scenario E where the production of $\tilde{\chi}^0_2$,
however, proceeds with sizable cross sections despite its large singlet
component.

In order to estimate the full potential of neutralino search
and to predict the chances for discriminating between NMSSM and MSSM at
a particular future electron-positron collider,
a detailed
Monte-Carlo simulation of the SM background as well as of the supersymmetric
processes has to be preformed. Moreover, it should include a discussion of
further relevant kinematical variables
as missing momenta, angular and energy distribution.
This goes well beyond the scope of this paper designed to
demonstrate fundamental differences in the neutralino sector
between the NMSSM and the minimal model .

\section{Conclusion}
Within the framework of the NMSSM, we analyzed production and decay
of neutralinos in seven typical scenarios that are compatible with
the experimental constraints from LEP and Tevatron. They
differ by the values for the gaugino mass parameter $M$ and for the
singlet vacuum expectation value $x$. In four scenarios, the LSP
is mainly a singlet with a
mass of 10 GeV well below the current mass bound for a MSSM
neutralino and therefore is mainly a singlet. In the other scenarios
where now the second lightest neutralino
has the largest singlet component
the LSP has a mass of 50 GeV.

Typical cross sections for the production of singlet-like neutralinos
at a 500-GeV electron-positron
linear collider reach values between 10 and 100 fb. Even if
the LSP is nearly massless, the next lightest
neutralino can already be so heavy that no visible neutralino
is produced at LEP2. Therefore, negative neutralino search at LEP2
is not expected to raise the
lower neutralino mass bound in the NMSSM.

In scenarios with light, singlet-like neutralinos there often exist
also light neutral Higgs bosons with masses below the MSSM bounds.
In typical NMSSM scenarios, the neutralino decay into Higgs bosons
may be dominant
and allow for a discrimination between MSSM
and NMSSM.
In the parameter range where
the decay into a Higgs boson and the LSP is kinematically
forbidden, the loop decay into a photon and the LSP is enhanced
compared to the three body decay into a fermion pair and the LSP, if
the neutralinos have significant singlet components.
For large singlet vacuum expectation values of about 1000 GeV,
the loop decay may dominate with branching ratios up to
\mbox{50 \%}.

If the lightest neutralino has a large singlet component, there are
good chances for distinguishing MSSM and NMSSM since it is the last link
of the decay chain and crucially determines the decay modes of the heavier
neutralinos. If, however, the second lightest neutralino is mainly a
singlet, the NMSSM can be verified only by its direct production and decay,
since the decay cascades of the heavier neutralinos then omit the singlet-like
neutralino.

Generally, neutralinos of the nonminimal model can manifest themselves
in a variety of different signatures.
Their identification can be facilitated by clear signatures with jets
or photons in the final state.
On the other hand, they could be invisible if they decay via Higgs bosons
mainly into the LSP. In any case
a careful analysis of
the SM background and Monte-Carlo studies considering the detector efficiency
are indispensable for the verification of minimal or nonminimal
supersymmetry.
\section*{Acknowledgements}
The authors would like to thank A.~Bartl for many helpful discussions.
Support by the Deutsche Forschungsgemeinschaft under contract
no.~FR 1064/2-1 is gratefully acknowledged.

\appendix
\section*{Appendix}
\section{Neutralino mixing}
The neutralino mass terms are contained in the following part of
the NMSSM Lagrangian
\begin{eqnarray}
{\cal L} & = & \frac{1}{\sqrt{2}}ig\lambda^3(v_1\psi_{H_1}^1-
v_2\psi_{H_2}^2)-\frac{1}{\sqrt{2}}ig'\lambda '
(v_1 \psi_{H_1}^1-v_2 \psi_{H_2}^2) \nonumber \\
& & - \frac{1}{2} M \lambda^3 \lambda^3
-\frac{1}{2} M' \lambda' \lambda' \nonumber \\
& & -\lambda x \psi_{H_1}^1 \psi_{H_2}^2
-\lambda v_1 \psi_{H_2}^2 \psi_N
-\lambda v_2 \psi_{H_1}^1 \psi_N
+kx\psi_N^2 \nonumber \\
& & \mbox{} \mbox{} + \mbox{h.c.}
\end{eqnarray}
where $\lambda ^3$, $\lambda '$, $\psi_{H_1}^1$ and $\psi_{H_2}^2$ are the
two component spinors of the supersymmetric partners
of the neutral gauge and Higgs bosons.
As a basis of the neutral gaugino-higgsino system we take
\begin{equation}
\label{basis}
(\psi^0)^T=(-i\lambda_{\gamma},-i\lambda_Z,\psi_H^a,\psi_H^b,
	    \psi_N)
\end{equation}
with the higgsino states
\begin{eqnarray}
\psi_H^a & = &
\psi_{H_1}^1 \cos \beta - \psi_{H_2}^2 \sin\beta
\nonumber \\
\psi_H^b & = &
\psi_{H_1}^1 \sin\beta + \psi_{H_2}^2 \cos\beta.
\end{eqnarray}
Then the mass term reads
\begin{equation}
{\cal L} = -\frac{1}{2} (\psi^0)^T Y \psi^0 + h.c.
\end{equation}
with the neutralino mass matrix
\begin{equation}
Y= \left(
\begin{array}{ccccc}
-M s^2_W-M' c^2_W &
(M'-M) s_W c_W &
0 & 0 & 0 \\
(M'-M) s_W c_W &
-M c^2_W-M' s^2_W &
m_Z & 0 & 0 \\
0 & m_Z &
-\lambda x \sin2\beta &
\lambda x \cos2\beta &
0 \\
0 & 0 &
\lambda x \cos2\beta &
\lambda x \sin2\beta &
\lambda v \\
0 & 0 & 0 &
\lambda v &
-2kx
\end{array}
\right).
\end{equation}
Here we introduced the abbreviations
\begin{equation}
s_W \equiv \sin\theta_W, \hspace*{1cm}
c_W \equiv \cos\theta_W, \hspace*{1cm}
v \equiv \sqrt{v_1^2+v_2^2}.
\end{equation}
In this paper we employ the usual gaugino mass relation
\begin{equation}
M'  =  \frac{5}{3} \frac{g_1}{g_2} M \simeq 0.5 M,
\end{equation}
with the couplings $g_i$ of the $U(1)$ and $SU(2)$ gauge groups.

In the limit $\lambda, \, k \longrightarrow 0$ with $\lambda x$ and
$kx$ fixed the neutralino mass matrix decouples, and the upper
$4 \times 4$ matrix
corresponds to that of the MSSM with $\mu = \lambda x $ \cite{fraas}.

The neutralino mass matrix can be diagonalized by a unitary
$5\times 5$ matrix N
\begin{equation}
m_{\tilde{\chi}^0_i}\delta_{ij} = N^\ast_{im}Y_{mn}N_{jn} ,
\end{equation}
where the $m_{\tilde{\chi}^0_i}$ are the mass eigenvalues of the neutralino
states
\begin{equation}
\chi_i^0 = N_{ij} \psi^0_j  \hspace*{1cm}
i,j = 1,\ldots , 5 \; .
\end{equation}
Finally, one obtains
the proper four-component neutralino mass eigenstates by defining
the Majorana spinors
\begin{equation}
\tilde{\chi}^0_i=\left( \begin{array}{c} \chi_i^0 \\
\bar{\chi}_i^0 \end{array} \right) \hspace*{1cm} i,j = 1,\ldots , 5 \; .
\end{equation}
\section{Higgs mixing}
Here we present the most important results for the Higgs sector.
Including radiative corrections from top and stop loops the
mass squared matrix for the neutral scalar Higgs bosons reads \cite{rad}
\begin{equation}
{\cal M}_S^2 = \left( \begin{array}{ccc}
{M_{11}^S}^2 & {M_{12}^S}^2 & {M_{13}^S}^2 \\
{M_{12}^S}^2 & {M_{22}^S}^2 & {M_{23}^S}^2 \\
{M_{13}^S}^2 & {M_{23}^S}^2 & {M_{33}^S}^2
\end{array} \right) + \delta {\cal M}_S^2
\end{equation}
with the matrix elements at tree level
\begin{eqnarray}
{M_{11}^S}^2 & = & \frac{(g^2+{g'}^2)v_1^2}{2}+
\lambda x(A_\lambda + kx) \tan\beta ,\\
{M_{12}^S}^2 & = & \frac{v_1v_2}{2}\left( 4\lambda ^2 -
g^2-{g'}^2 \right) -\lambda x(A_\lambda+kx), \\
{M_{13}^S}^2 & = & 2 \lambda ^2 v_1 x -2\lambda kxv_2 -
\lambda A_\lambda v_2, \\
{M_{22}^S}^2 & = & \frac{(g^2+{g'}^2)v_2^2}{2}+
\lambda x(A_\lambda + kx) \cot\beta ,\\
{M_{23}^S}^2 & = & 2 \lambda ^2 v_2 x -2\lambda kxv_1 -
\lambda A_\lambda v_1, \\
{M_{33}^S}^2 & = & 4k^2x^2-kA_kx + \frac{\lambda A_\lambda v_1 v_2}{x},
\end{eqnarray}
and the radiative corrections
\begin{equation}
\delta {\cal M}_S^2 =
\left( \begin{array}{ccc}
\Delta_{11}^2 & \Delta_{12}^2 & \Delta_{13}^2 \\
\Delta_{12}^2 & \Delta_{22}^2 & \Delta_{23}^2 \\
\Delta_{13}^2 & \Delta_{23}^2 & \Delta_{33}^2
\end{array} \right) +
\left( \begin{array}{ccc}
\tan\beta & -1 & -\frac{v_2}{x} \\
-1 & \cot\beta & -\frac{v_1}{x} \\
-\frac{v_2}{x} & -\frac{v_1}{x} & \frac{v_1v_2}{x^2}
\end{array} \right) \Delta^2
\end{equation}
with
\begin{eqnarray}
\Delta^2 & = & \frac{3}{16 \pi^2} h_t^2 \lambda x A_t f(m^2_{\tilde{t}_1},
m^2_{\tilde{t}_2}), \\
\Delta_{11}^2 & = & \frac{3}{8 \pi^2} h_t^4 v_2^2 \lambda^2 x^2
\left( \frac{A_t+\lambda x \cot\beta}{m^2_{\tilde{t}_2}-
m^2_{\tilde{t}_1}} \right) ^2 g(m^2_{\tilde{t}_1},m^2_{\tilde{t}_2}), \\
\Delta_{12}^2 & = & \frac{3}{8 \pi^2} h_t^4 v_2^2 \lambda x
\left( \frac{A_t+\lambda x \cot\beta}{m^2_{\tilde{t}_2}-
m^2_{\tilde{t}_1}} \right) \nonumber \\
& & \times \left( \log \left( \frac{m^2_{\tilde{t}_2}}{m^2_{\tilde{t}_1}}
\right) + \frac{A_t (A_t+\lambda x \cot\beta)}{m^2_{\tilde{t}_2}-
m^2_{\tilde{t}_1}} g(m^2_{\tilde{t}_1},m^2_{\tilde{t}_2}) \right), \\
\Delta_{13}^2  & = & \frac{3}{8 \pi^2} h_t^4 v_2^2 \lambda^2 v_1 x
\left( \frac{A_t+\lambda x \cot\beta}{m^2_{\tilde{t}_2}-
m^2_{\tilde{t}_1}} \right) ^2 g(m^2_{\tilde{t}_1},m^2_{\tilde{t}_2})
\nonumber \\ & &
-\frac{3}{8 \pi^2} h_t^4 \lambda^2 v_1 x f(m^2_{\tilde{t}_1},m^2_{\tilde{t}_2})
,\\
\Delta_{22}^2 & = & \frac{3}{8 \pi^2} h_t^4 v_2^2
\left( \log \left( \frac{m^2_{\tilde{t}_1}m^2_{\tilde{t}_2}}{m_t^4} \right)
+\frac{2A_t (A_t+\lambda x \cot\beta)}{m^2_{\tilde{t}_2}-
m^2_{\tilde{t}_1}}
\log \left( \frac{m^2_{\tilde{t}_2}}{m^2_{\tilde{t}_1}}
\right) \right) \nonumber \\ & &
+\frac{3}{8 \pi^2} h_t^4 v_2^2 \left(
\frac{A_t (A_t+\lambda x \cot\beta)}{m^2_{\tilde{t}_2}-
m^2_{\tilde{t}_1}} \right) ^2 g(m^2_{\tilde{t}_1},m^2_{\tilde{t}_2}), \\
\Delta_{23}^2 & = & \frac{3}{8 \pi^2} h_t^4 v_2^2 \lambda v_1
\left( \frac{A_t+\lambda x \cot\beta}{m^2_{\tilde{t}_2}-
m^2_{\tilde{t}_1}} \right) \nonumber \\
& & \times \left( \log \left( \frac{m^2_{\tilde{t}_2}}{m^2_{\tilde{t}_1}}
\right) + \frac{A_t (A_t+\lambda x \cot\beta)}{m^2_{\tilde{t}_2}-
m^2_{\tilde{t}_1}} g(m^2_{\tilde{t}_1},m^2_{\tilde{t}_2}) \right), \\
\Delta_{33}^2 & = & \frac{3}{8 \pi^2} h_t^4 v_2^2 \lambda ^2 v_1^2
\left( \frac{A_t+\lambda x \cot\beta}{m^2_{\tilde{t}_2}-
m^2_{\tilde{t}_1}} \right) ^2 g(m^2_{\tilde{t}_1},m^2_{\tilde{t}_2}).
\end{eqnarray}
Here $m_{\tilde{t}_1}$ and $m_{\tilde{t}_2}$ are the stop masses,
$A_t$ is the mass parameter in the soft symmetry breaking potential
eq.~(\ref{soft}) connected with the top quark. The functions $f$ and $g$
are defined as
\begin{eqnarray}
f(m^2_{\tilde{t}_1},m^2_{\tilde{t}_2}) & = &
\frac{1}{m^2_{\tilde{t}_2}-m^2_{\tilde{t}_1}}
\left( m^2_{\tilde{t}_1} \log \left(
\frac{m^2_{\tilde{t}_1}}{\mu^2} \right)-
m^2_{\tilde{t}_2} \log \left(
\frac{m^2_{\tilde{t}_2}}{\mu^2} \right)
-m^2_{\tilde{t}_1}+m^2_{\tilde{t}_2} \right) , \; \; \; \; \; \; \; \\
g(m^2_{\tilde{t}_1},m^2_{\tilde{t}_2}) & = &
\frac{-1}{m^2_{\tilde{t}_2}-m^2_{\tilde{t}_1}}
\left( (m^2_{\tilde{t}_1}+m^2_{\tilde{t}_2}) \log \left(
\frac{m^2_{\tilde{t}_2}}{m^2_{\tilde{t}_1}} \right)
+2(m^2_{\tilde{t}_1}-m^2_{\tilde{t}_2}) \right)
\end{eqnarray}
with the $\overline{\mbox{MS}}$ renormalization scale $\mu$.

For the pseudoscalar Higgs bosons one obtains the mass squared matrix
\begin{equation}
{\cal M}_P^2 = \left( \begin{array}{ccc}
{M_{11}^P}^2 & {M_{12}^P}^2 & {M_{13}^P}^2 \\
{M_{12}^P}^2 & {M_{22}^P}^2 & {M_{23}^P}^2 \\
{M_{13}^P}^2 & {M_{23}^P}^2 & {M_{33}^P}^2
\end{array} \right) + \delta {\cal M}_P^2,
\end{equation}
where the tree level matrix elements are
\begin{eqnarray}
{M_{11}^P}^2 & = &
\lambda x(A_\lambda + kx) \tan\beta ,\\
{M_{12}^P}^2 & = &
\lambda x(A_\lambda + kx)  ,\\
{M_{13}^P}^2 & = &  \lambda v_2 ( A_\lambda -2kx), \\
{M_{22}^P}^2 & = &
\lambda x(A_\lambda + kx) \cot\beta ,\\
{M_{23}^P}^2 & = &  \lambda v_1 ( A_\lambda -2kx), \\
{M_{33}^P}^2 & = & \lambda A_\lambda \frac{v_1v_2}{x} + 4\lambda k v_1 v_2
+3kA_kx ,
\end{eqnarray}
and the radiative corrections can be written as
\begin{equation}
\delta {\cal M}_P^2 =
\left( \begin{array}{ccc}
\tan\beta & 1 & \frac{v_2}{x} \\
1 & \cot\beta & \frac{v_1}{x} \\
 \frac{v_2}{x} & \frac{v_1}{x} & \frac{v_1v_2}{x^2}
\end{array} \right) \Delta^2  .
\end{equation}
One eigenvalue of the mass matrix for the pseudoscalar Higgs bosons
corresponds to an unphysical massless Goldstone mode.

The mass eigenstates of the scalar Higgs bosons are denoted by
$S_i$ ($i=1,2,3$), ($m_{S_1} \le m_{S_2} \le m_{S_3}$) and those
of the physical pseudoscalars by
$P_j$ ($j=1,2$), ($m_{P_1} \le m_{P_2}$).
They follow from the Higgs fields $H_1^0$, $H_2^0$ and $N$
by transformation with the diagonalization matrices $U^S$ and $U^P$
\begin{eqnarray}
\label{higgstranss}
\left( \begin{array}{c} S_1 \\ S_2 \\ S_3 \end{array} \right) & = &
\sqrt{2} {U^S}  \left[
\left( \begin{array}{c} \mbox{Re} H_1^0 \\
\mbox{Re} H_2^0 \\ \mbox{Re} N \end{array} \right) -
\left( \begin{array}{c} v_1 \\ v_2 \\ x \end{array} \right) \right] \; ,
\\ & & \nonumber \\
\left( \begin{array}{c} P_1 \\ P_2   \end{array} \right) & = &
\sqrt{2}
U^P
\left( \begin{array}{c} \mbox{Im} H_1^0 \\
\mbox{Im} H_2^0 \\ \mbox{Im} N \end{array} \right) \; .
\label{higgstransp}
\end{eqnarray}
Since we omit the unphysical Goldstone bosons, $U^P$ is a $3 \times 2$
matrix transforming the imaginary parts of the Higgs fields in the
Lagrangian to the physical pseudoscalar Higgs bosons.

\section{Cross sections and decay widths}
For the analytical formulae for neutralino production and decay we use
the following notation:
\begin{enumerate}
\item the parameters in the couplings between a $Z$ boson and two fermions
are
\begin{equation}
L_f=T_{3f}-e_f\sin^2\theta_W, \hspace{0.5cm} R_f=-e_f\sin^2\theta_W,
\end{equation}
where $e_f$ and $T_{3f}$ denote the charge and the isospin of the fermion,
respectively;
\item the parameters in the coupling between a $Z$ boson and two neutralinos
are
\begin{eqnarray}
O_{ij}^{''L} & = & -\frac{1}{2} (N_{i3}N_{j3}^\ast - N_{i4}N_{j4}^\ast)
\cos 2 \beta
-\frac{1}{2} (N_{i3}N_{j4}^\ast - N_{i4}N_{j3}^\ast)
\sin 2 \beta , \\
O_{ij}^{''R} & = & -O_{ij}^{''L\ast},
\end{eqnarray}
where $N_{ij}$ denote the mixing components of the neutralinos in the
basis (\ref{basis});
\item the parameters in the couplings between a scalar Higgs boson and
two neutralinos are
\begin{eqnarray}
Q^{L''}_{aij} & = &
\frac{1}{2} \left[
(U^S_{a1}\cos\beta+U^S_{a2}\sin\beta)
\left(\frac{g}{c_W}
(N_{i2}N_{j3}^{\ast}+N_{j2}N_{i3}^{\ast}) \right. \right.
\nonumber \\ & & \left. \left.
+\sqrt{2}\lambda(N_{i5}N_{j4}^{\ast}+N_{j5}N_{i4}^{\ast})
\right) \right.
\nonumber \\ & & \left.
+(U^S_{a1}\sin\beta-U^S_{a2}\cos\beta)
\left(\frac{g}{c_W}
(N_{i2}N_{j4}^{\ast}+N_{j2}N_{i4}^{\ast}) \right. \right.
\nonumber \\ & & \left. \left.
-\sqrt{2}\lambda (N_{i5}N_{j3}^{\ast}+N_{j5}N_{i3}^{\ast})
\right) \right]
\nonumber \\ & &
-\sqrt{2}kU^S_{a3}(N_{i5}N_{j5}^{\ast}+N_{j5}N_{i5}^{\ast}) ,
\\
Q^{R''}_{aij} & = & Q^{L''\ast}_{aij} ,
\end{eqnarray}
where $U^S$ is the diagonalization matrix of the scalar Higgs sector
in eq.~(\ref{higgstranss});
\item the parameters in the couplings between a pseudoscalar Higgs boson
and two neutralinos are
\begin{eqnarray}
\label{higgsneuneu}
R^{L''}_{\alpha ij} & = &
-\frac{1}{2} \left[
(U^P_{\alpha 1}\cos\beta+U^P_{\alpha 2}\sin\beta)
\left(\frac{g}{c_W}(N_{i2}N_{j3}^{\ast}+N_{j2}N_{i3}^{\ast}) \right. \right.
\nonumber \\ & & \left. \left.
-\sqrt{2}\lambda(N_{i5}N_{j4}^{\ast}+N_{j5}N_{i4}^{\ast})
\right) \right.
\nonumber \\ & & \left.
+(U^P_{\alpha 1}\sin\beta-U^P_{\alpha 2}\cos\beta)
\left(\frac{g}{c_W}
(N_{i2}N_{j4}^{\ast}+N_{j2}N_{i4}^{\ast}) \right. \right.
\nonumber \\ & & \left. \left.
+\sqrt{2}\lambda(N_{i5}N_{j3}^{\ast}+N_{j5}N_{i3}^{\ast})
\right) \right] ,
\nonumber \\ & &
-\sqrt{2}kU^P_{\alpha 3}(N_{i5}N_{j5}^{\ast}+N_{j5}N_{i5}^{\ast})
\\
R^{R''}_{\alpha ij} & = & -R^{L''\ast}_{\alpha ij},
\end{eqnarray}
where $U^P$ is the diagonalization matrix of the pseudoscalar Higgs sector
in eq.~(\ref{higgstransp});
\item the parameters in the coupling between a neutralino, a scalar lepton or
scalar quark and
a lepton or quark are
\begin{eqnarray}
f^L_{fi} & = & -\sqrt{2} \left[ \frac{1}{\cos\theta_W}L_f N_{i2}
-\frac{1}{\sin\theta_W}R_f N_{i1} \right] \\
f^R_{fi} & = & -\sqrt{2}e_f \sin\theta_W [\tan\theta_W N_{i2}^\ast-
N_{i1}^\ast ];
\end{eqnarray}
\item the $Z$ propagator
\begin{eqnarray}
D_Z (x) & = & (x-m_Z^2+im_Z\Gamma_Z)^{-1} ;
\end{eqnarray}
\item the triangle function
\begin{equation}
\lambda(a,b,c) = a^2+b^2+c^2-2ab-2ac-2bc.
\end{equation}
\end{enumerate}

Then one finds for the cross sections of the process
$e^+e^- \longrightarrow \tilde{\chi}^0_i \tilde{\chi}^0_j$ \cite{bartlneuprod}
\begin{equation}
\sigma_{\mbox{tot}}  =  \frac{1}{2} \left( \sigma_Z + \sigma_{\tilde{e}}
+\sigma_{Z\tilde{e}} \right) \left( 2-\delta_{ij} \right).
\end{equation}
The particular terms arise from $Z$ and selectron exchange and
their interference
\begin{eqnarray}
\sigma_Z & = & \frac{g^4}{4\pi \cos^4 \theta_W} |D_Z(s)|^2
\frac{q}{\sqrt{s}} |O_{ij}^{''L}|^2 \left( |L_e|^2+|R_e|^2 \right)
\left[ E_i E_j + \frac{1}{3} q^2 - m_i m_j \right] , \\
\sigma_{\tilde{e}} & = & \frac{g^4}{16 \pi} \frac{q}{s\sqrt{s}}
\left\{ |f_{ei}^L|^2|f_{ej}^L|^2 \left[ \frac{E_iE_j-sd_L+q^2}{sd_l-q^2}
+2+\frac{\sqrt{s}}{2q} \left(1-2d_L-\frac{m_im_j}{sd_L} \right) \right.
\right. \nonumber \\ & & \left. \left. \times
\ln \left| \frac{d_L+q/\sqrt{s}}{d_L-q/\sqrt{s}}\right| \right]
+ (L \leftrightarrow R)
\right\}, \\
\sigma_{Z\tilde{e}} & = & -\frac{g^4}{8\pi \cos^2\theta_W}
\frac{q}{\sqrt{s}} \mbox{Re} \left( D_Z(s) \right) O_{ij}^{''L}
\nonumber \\ & & \times \left\{ L_ef^L_{ei}f^L_{ej}
\left[ \frac{1}{q\sqrt{s}} \left( E_iE_j-sd_L(1-d_L)-m_im_j\right)
\ln \left| \frac{d_L+q/\sqrt{s}}{d_L-q/\sqrt{s}}\right| +2(1-d_L)
\right] \right.
\nonumber \\ & &
- (L \leftrightarrow R) \Bigg\}
\end{eqnarray}
with
\begin{equation}
d_{L,R}= \frac{1}{2s} \left( s+2m_{\tilde{e}_{L,R}}^2-m_i^2-m_j^2 \right).
\end{equation}
Here, $q$ is the momentum of $\tilde{\chi}^0_i$ in the $e^+e^-$
center-of-mass system
\begin{equation}
q= \frac{1}{2 \sqrt{s}} \sqrt{\lambda( s,m_i^2,m_j^2)},
\end{equation}
and $E_i = \sqrt{q^2+m_i^2}$.

\begin{figure}
\begin{center}
\begin{picture}(16,20)
\put(1,13){\includegraphics{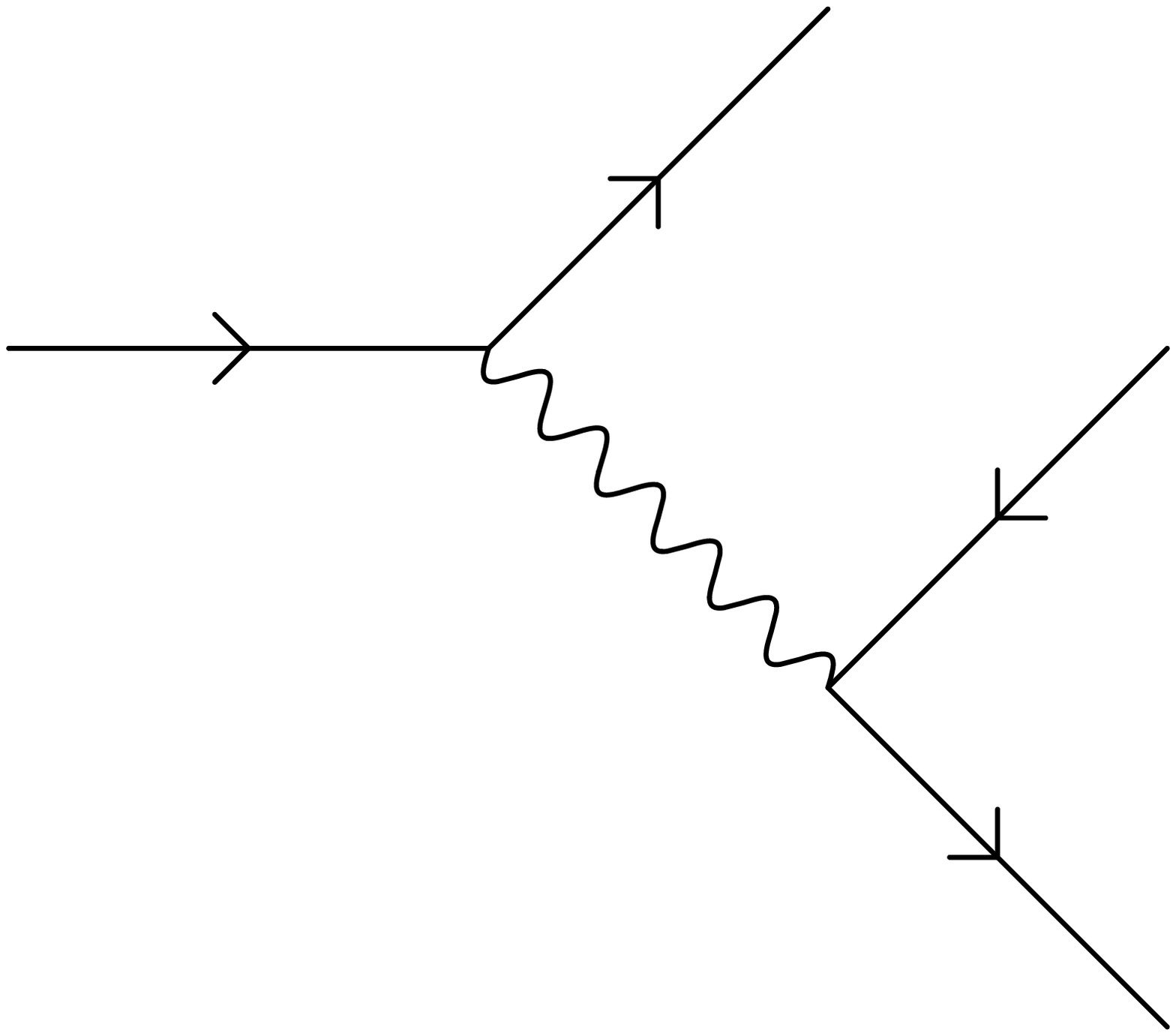}}
\put(9,15){\includegraphics{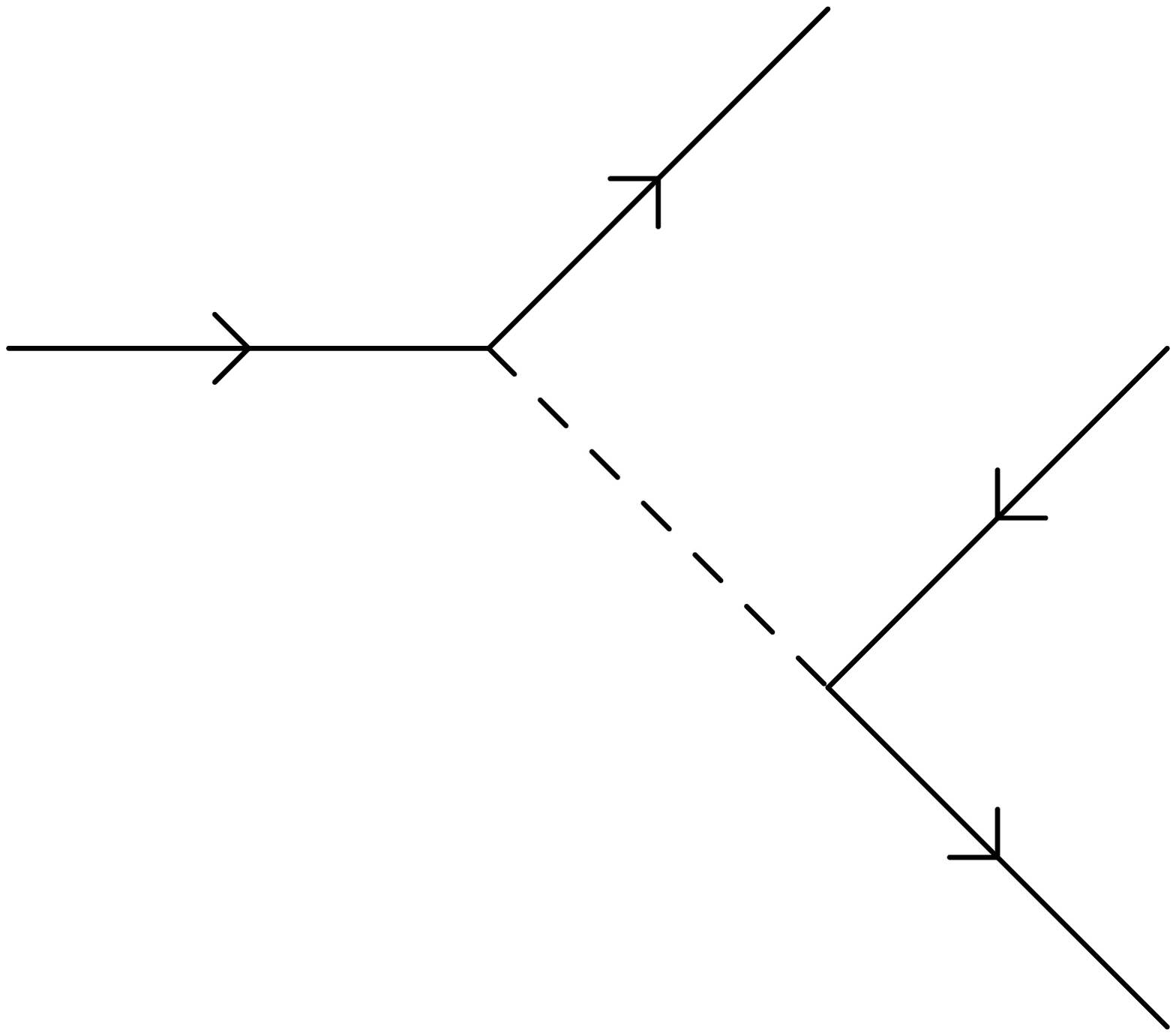}}
\put(9,10.5){\includegraphics{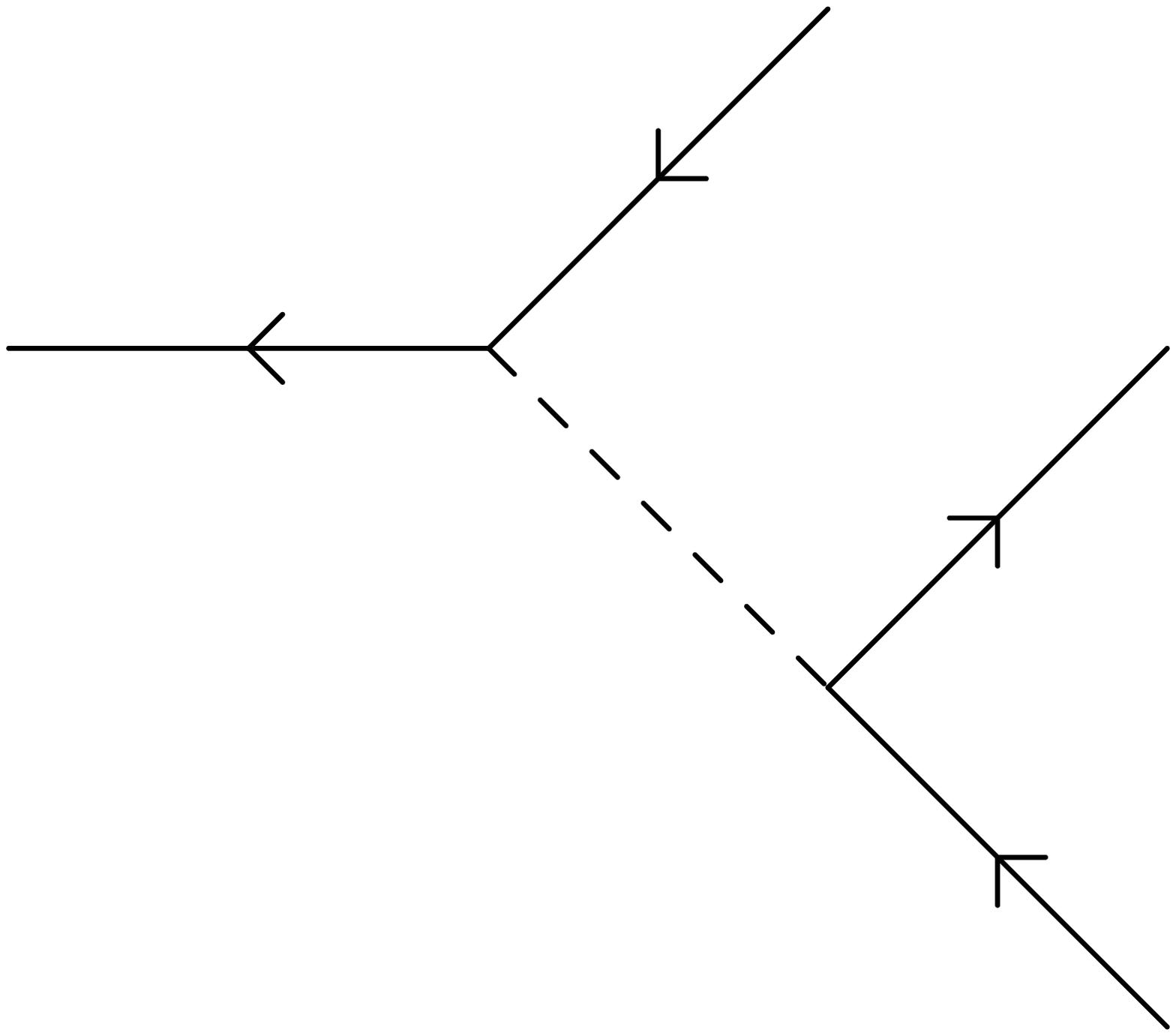}}
\put(0.8,5.3){\includegraphics{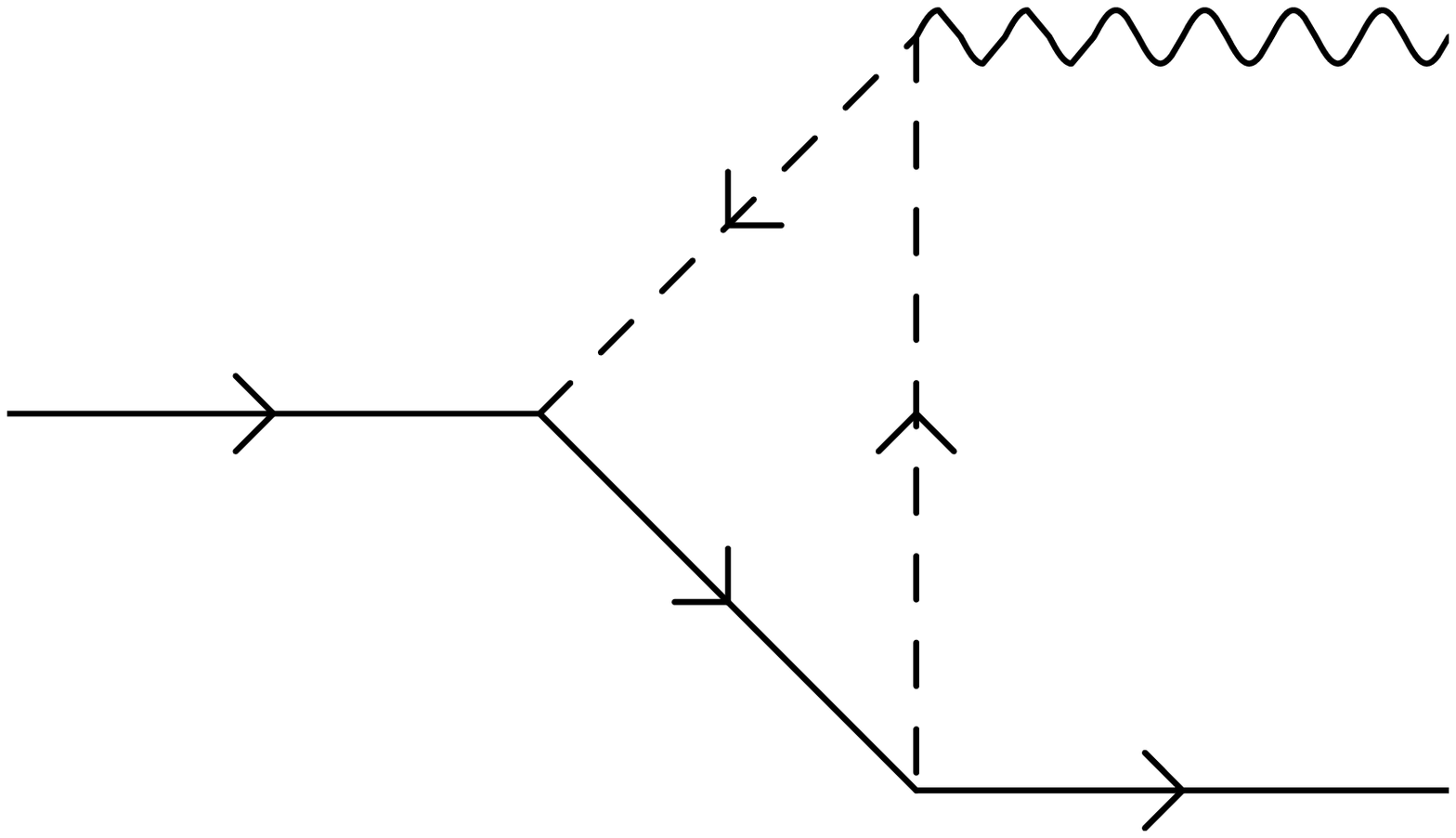}}
\put(8.0,5.3){\includegraphics{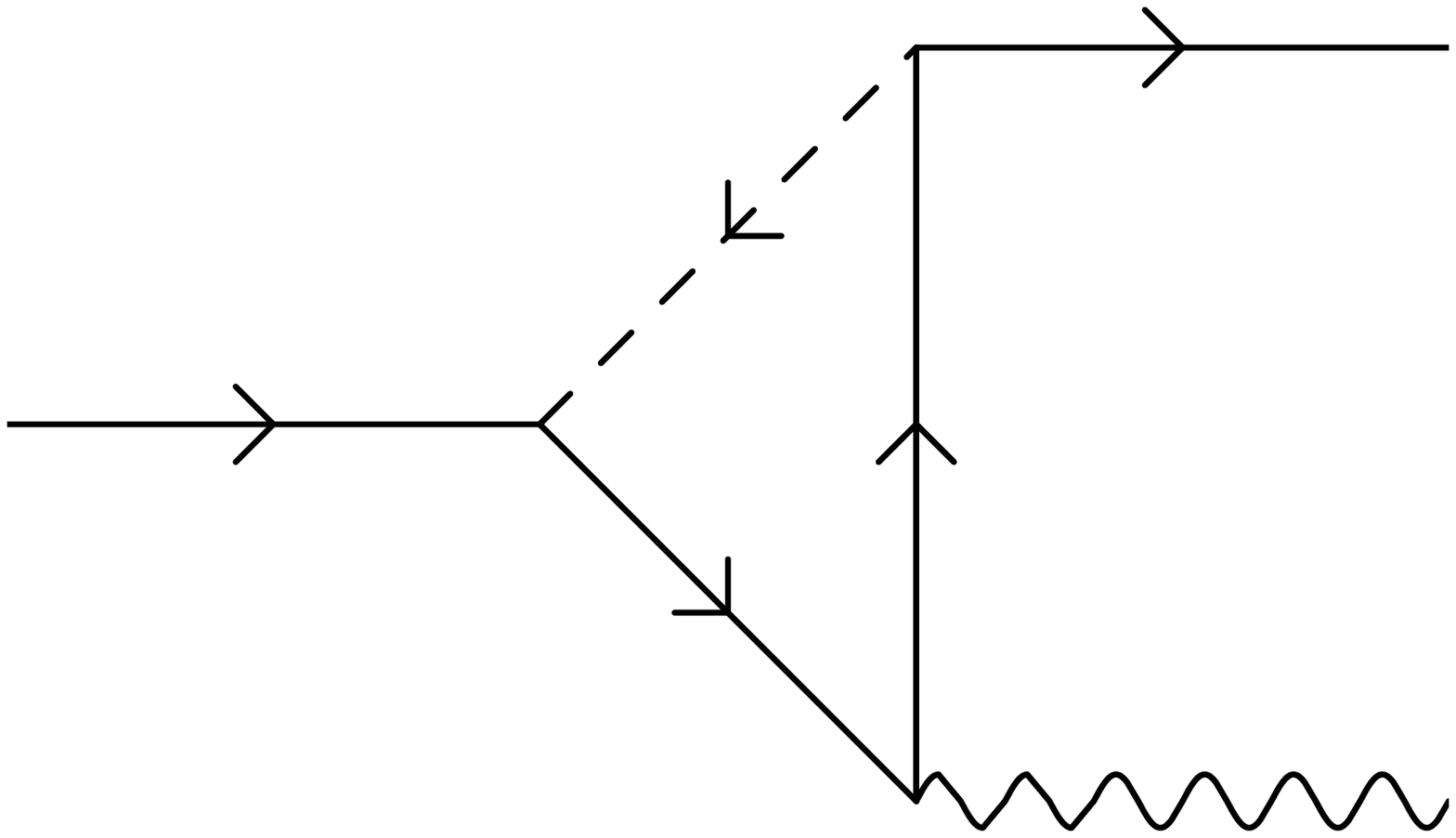}}
\put(1,-1.5){\includegraphics{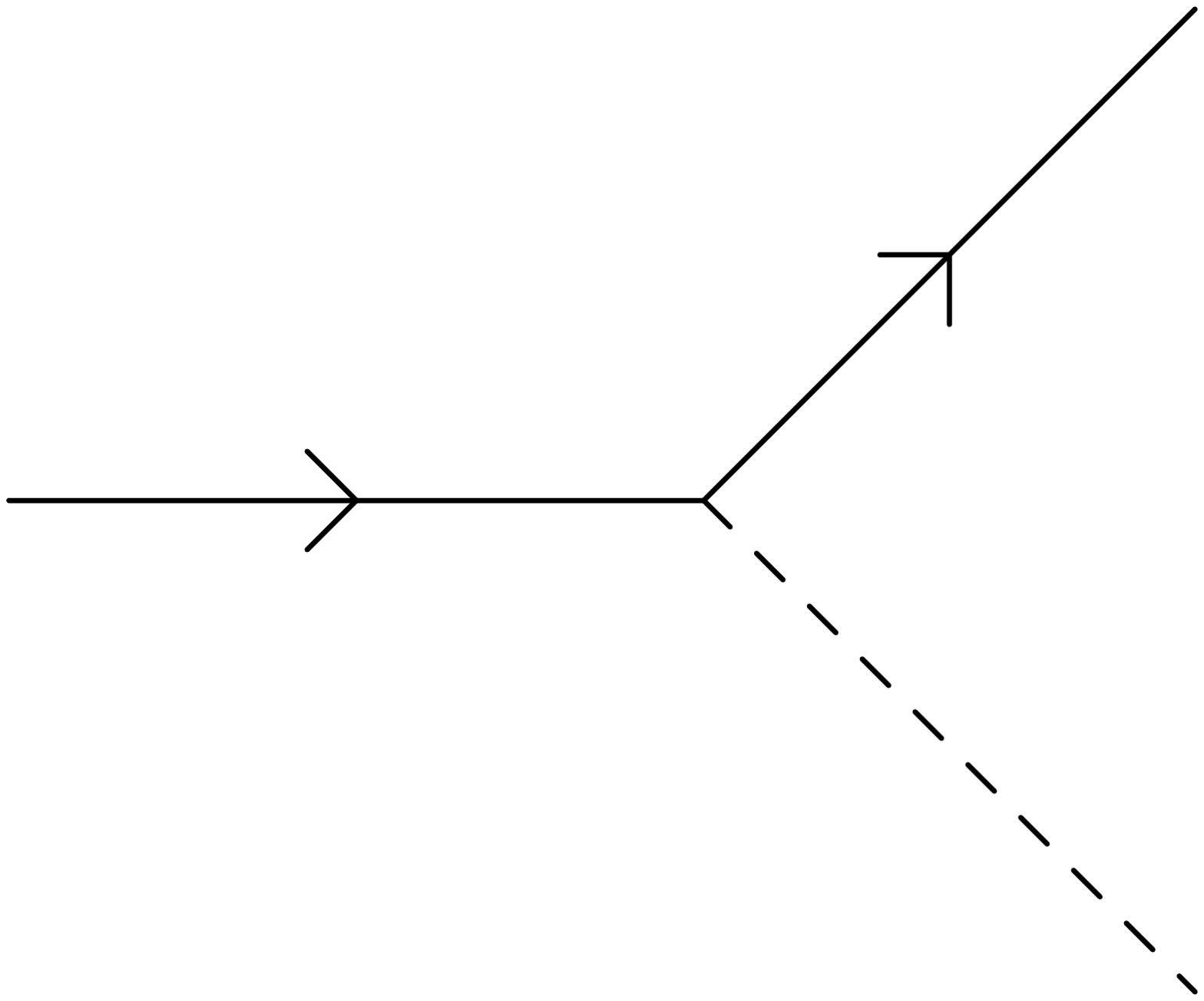}}
\put(1.0,19){$\bullet \;
\tilde{\chi}^0_i \longrightarrow \tilde{\chi}^0_j f \bar{f}$:}
\put(1.0,12.0){$\bullet \;
\tilde{\chi}^0_i \longrightarrow \tilde{\chi}^0_j \gamma$:}
\put(1.0,4.5){$\bullet \;
\tilde{\chi}^0_i \longrightarrow \tilde{\chi}^0_j S_a/P_\alpha $:}
\put(1.5,6.6){+ corresponding graphs with clockwise circulating particles
in the loop}
\put(1.5,6.1){+ 8 other graphs with MSSM couplings (for details
see \cite{wyler})}
\put(1.0,17){$\tilde{\chi}^0_i$}
\put(4.7,18.4){$\tilde{\chi}^0_j$}
\put(4.0,16.6){$Z$}
\put(5.9,17.2){$\bar{f}$}
\put(5.9,14.5){$f$}
\put(9.0,19){$\tilde{\chi}^0_i$}
\put(12.7,20.4){$f$}
\put(12.0,18.6){$\tilde{f}_{L,R}$}
\put(13.9,19.2){$\bar{f}$}
\put(13.9,16.5){$\tilde{\chi}^0_j$}
\put(9.0,14.5){$\tilde{\chi}^0_i$}
\put(12.7,15.9){$\bar{f}$}
\put(12.0,14.1){$\tilde{f}_{L,R}$}
\put(13.9,14.7){$f$}
\put(13.9,12.0){$\tilde{\chi}^0_j$}
\put(1.0,9.4){$\tilde{\chi}^0_i$}
\put(3.5,8.4){$\tilde{\chi}^+_k$}
\put(2.7,10.3){$H^+$,$G^+$}
\put(5.0,9.3){$H^+$,$G^+$}
\put(6.7,10.7){$\gamma$}
\put(6.7,8.1){$\tilde{\chi}^0_j$}
\put(8.2,9.4){$\tilde{\chi}^0_i$}
\put(10.7,8.4){$\tilde{\chi}^+_k$}
\put(9.9,10.3){$H^+$,$G^+$}
\put(12.2,9.4){$\tilde{\chi}^+_k$}
\put(13.9,8.1){$\gamma$}
\put(13.9,10.7){$\tilde{\chi}^0_j$}
\put(1.0,2.0){$\tilde{\chi}^0_i$}
\put(5.9,3.8){$\tilde{\chi}^0_j$}
\put(5.9,-0.2){$S_a/P_\alpha $}
\end{picture}
\end{center}
\caption{Feynman graphs for the considered neutralino decays.}
\label{zerfey}
\end{figure}

The decay width for the three body decay of a neutralino as shown
in Fig.~\ref{zerfey} can be written as
\begin{equation}
\Gamma (\tilde{\chi}_i^0 \rightarrow \tilde{\chi}_j^0 + f + \bar{f})
=  N_c \frac{g^4}{(2\pi)^3 64 |m_i|^3} \int d\bar{s} d\bar{t}
(W_s+W_t+W_u+W_{tu}+W_{st}+W_{su})
\end{equation}
with
\begin{eqnarray}
W_s & = & |D_Z(\bar{s})|^2 \frac{4|O_{ij}^{''L}|^2 (L_f^2+R_f^2)}
{\cos^4\theta_W} \nonumber \\ & & \times
\left[ (m_i^2-\bar{t})(\bar{t}-m_j^2)+(m_i^2-\bar{u})(\bar{u}-m_j^2)
+2m_im_j\bar{s} \right], \\
W_t & = & |f_{fi}^L|^2 |f_{fj}^L|^2 \frac{(m_i^2-\bar{t})(\bar{t}-m_j^2)}
{(\bar{t}-m_{\tilde{f}_L}^2)^2} + (L \leftrightarrow R), \\
W_u & = & W_t (\bar{t} \leftrightarrow \bar{u}), \\
W_{tu} & = & |f_{fi}^L|^2 |f_{fj}^L|^2 \frac{2m_im_j\bar{s}}
{(\bar{t}-m_{\tilde{f}_L}^2)(\bar{u}-m_{\tilde{f}_L}^2)}
+ (L \leftrightarrow R), \\
W_{st} & = & \frac{4 \mbox{Re} (D_Z(\bar{s}))}{(\bar{t}-m_{\tilde{f}_L}^2)}
\frac{f_{fi}^lf_{fj}^LO_{ij}^{''L}L_f}{\cos^2\theta_W} \nonumber \\ & &
\times \left[ (m_i^2-\bar{t})(\bar{t}-m_j^2)+m_im_j\bar{s} \right]
+ (L \leftrightarrow R) ,\\
W_{su} & = & W_{st} (\bar{t} \leftrightarrow \bar{u}).
\end{eqnarray}
Here the integration
bounds are $\bar{s}_{\mbox{\scriptsize min}}=0$,
$\bar{s}_{\mbox{\scriptsize max}}
=(|m_i|-|m_j|)^2$, $\bar{t}_{\mbox{\scriptsize max,min}}=
\frac{1}{2}\left( m_i^2+m_j^2-\bar{s} \right.$ $\left. \pm
\sqrt{\lambda(m_i^2,m_j^2,\bar{s})}\right)$,
and $\bar{u}=m_i^2+m_j^2-\bar{s}-\bar{t}$.
The factor $N_c$ has the value 1 for the decay into leptons and 3 for the
decay into a quark pair.

Further we computed in this paper the decay width for the loop decay
into a lighter neutralino and a photon. Due to its lengthy structure
we do not state the analytical formula. It can be found in ref.~\cite{wyler}
for the MSSM, for the NMSSM one has to replace the parameters
$Q_{ij}^{'L}$ and $Q_{ij}^{'R}$ in the $\tilde{\chi}^0
\tilde{\chi}^\pm H^\mp$ couplings by
\begin{eqnarray}
Q^{'L}_{ij} = & g\cos\beta & \bigg[
\left( -N_{i3}\sin\beta+N_{i4}\cos\beta\right) V_{j1}
\nonumber \\ & &
+\left. \frac{1}{\sqrt{2}} \left( 2s_WN_{i1}+
(c_W-\frac{s_W^2}{c_W})N_{i2} \right) V_{j2} \right]
\nonumber \\ & -\lambda^{\ast}\sin\beta & N_{i5}V_{j2} ,\\
Q^{'R}_{ij} = & g\sin\beta & \bigg[
\left( N_{i3}\cos\beta+N_{i4}\sin\beta\right) U_{j1}
\nonumber \\ & &
-\left. \frac{1}{\sqrt{2}} \left( 2s_WN_{i1}+
(c_W-\frac{s_W^2}{c_W})N_{i2} \right) U_{j2} \right]
\nonumber \\ & -\lambda^{\ast}\cos\beta & N_{i5}U_{j2} .
\label{neucharhiggs}
\end{eqnarray}
For details of the NMSSM couplings see ref.~\cite{franke3}.

The decay widths for the neutralino decay into a lighter neutralino
and a neutral scalar or pseudoscalar Higgs boson read
\begin{eqnarray}
\Gamma (\tilde{\chi}^0_i \rightarrow  \tilde{\chi}^0_j + S_a) & = &
\frac{\sqrt{\lambda(m_i^2,m_j^2,m_a^2)}}{16\pi |m_i|^3}
{Q_{aij}^{''L}}^2 \left[ (m_i^2+m_j^2-m_a^2)+2 m_i m_j \right] ,
\\
\Gamma (\tilde{\chi}^0_i \rightarrow  \tilde{\chi}^0_j + P_\alpha ) & = &
\frac{\sqrt{\lambda(m_i^2,m_j^2,m_\alpha ^2)}}{16\pi |m_i|^3}
{R_{\alpha ij}^{''L}}^2 \left[ (m_i^2+m_j^2-m_\alpha ^2)-2 m_i m_j \right].
\end{eqnarray}

For the subsequent decays of the
produced light Higgs bosons
one obtains the following decay widths for
\begin{enumerate}
\item the decay of a scalar Higgs boson into a pair of up type fermions
\begin{eqnarray}
\Gamma (S_a \rightarrow f \bar{f}) & = &
N_c \frac{g^2 m_f^2}{32 \pi m_W^2 \sin^2\beta} m_a {U^S_{a2}}^2
\left( 1-\frac{4m_f^2}{m_a^2} \right) ^{3/2} ;
\end{eqnarray}
\item the decay of a scalar Higgs boson into a pair of down type fermions
\begin{eqnarray}
\Gamma (S_a \rightarrow f \bar{f}) & = &
N_c \frac{g^2 m_f^2}{32 \pi m_W^2 \cos^2\beta} m_a {U^S_{a1}}^2
\left( 1-\frac{4m_f^2}{m_a^2} \right) ^{3/2} ;
\end{eqnarray}
\item the decay of a pseudoscalar Higgs boson into a pair of up type fermions
\begin{eqnarray}
\Gamma (P_\alpha \rightarrow f \bar{f}) & = &
N_c \frac{g^2 m_f^2}{32 \pi m_W^2 \sin^2\beta} m_\alpha {U^P_{\alpha 2}}^2
\left( 1-\frac{4m_f^2}{m_a^2} \right) ^{1/2} ;
\end{eqnarray}
\item the decay of a pseudoscalar Higgs boson into a pair of down type fermions
\begin{eqnarray}
\Gamma (P_\alpha \rightarrow f \bar{f}) & = &
N_c \frac{g^2 m_f^2}{32 \pi m_W^2 \cos^2\beta} m_\alpha {U^P_{\alpha 1}}^2
\left( 1-\frac{4m_f^2}{m_a^2} \right) ^{1/2} ;
\end{eqnarray}
\item the decay of a scalar or pseudoscalar Higgs boson into two neutralinos
\begin{eqnarray} \hspace*{-0.5cm}
\Gamma (S_a \rightarrow \tilde{\chi}^0_i \tilde{\chi}^0_j) & = &
\frac{\sqrt{\lambda(m_a^2,m_i^2,m_j^2)}}{8 \pi m_a^3 (1+\delta_{ij})}
{Q_{aij}^{''L}}^2 \left[ \left( m_a^2-m_i^2-m_j^2 \right) -2 m_im_j \right]
, \; \; \; \;
\\ \hspace*{-0.5cm}
\Gamma (P_\alpha \rightarrow \tilde{\chi}^0_i \tilde{\chi}^0_j) & = &
\frac{\sqrt{\lambda(m_\alpha^2,m_i^2,m_j^2)}}{8 \pi m_\alpha^3 (1+\delta_{ij})}
{R_{\alpha ij}^{''L}}^2 \left[ \left( m_\alpha^2-m_i^2-m_j^2 \right) +2 m_im_j
\right]
; \; \; \; \;
\end{eqnarray}
\item the decay of a scalar Higgs boson into two light pseudoscalar
Higgs particles
\begin{eqnarray}
\Gamma (S_a \rightarrow P_\beta P_\gamma ) & = &
\frac{\sqrt{\lambda(m_a^2,m_\beta^2,m_\gamma^2)}}{16 \pi m_a^3
(1+\delta_{\beta \gamma})}
g^2_{S_aP_\beta P_\gamma}  .
\end{eqnarray}
The couplings $g_{S_aP_\beta P_\gamma }$ between one scalar and
two pseudoscalar
Higgs bosons can be found in ref.~\cite{franke3}.
\end{enumerate}

\newpage
\begin{table}
\begin{center}
\begin{tabular}{||l|l|cccccc||}
\hline
\hline
\multicolumn{8}{||c||}{} \\
\multicolumn{8}{||c||}{Scenario A} \\
\multicolumn{8}{||c||}{$M=65$ GeV, $x=1000$ GeV, $\lambda=0.4$, $k=0.001$,
$\tan\beta=2$} \\ \multicolumn{8}{||c||}{} \\ \hline
& & Mass & Photino & Zino & Higgsino A & Higgsino B & Singlet \\
& & [GeV] & & & & & \\
& $\tilde{\chi}^0_1$ & $-8$ & $-.141$ & $0.329$ & $0.157$ & $-.079$ &
$0.917$ \\
& $\tilde{\chi}^0_2$ & $-28$ & $-.712$ & $0.602$ & $0.086$ & $0.125$ &
$-.330$ \\
Neutralinos & $\tilde{\chi}^0_3$ & $-54$ & $0.688$ & $0.687$ &
$0.133$ & $0.114$ & $-.125$ \\
& $\tilde{\chi}^0_4$ & $412$ & $-.002$ & $0.061$ & $0.314$ & $-.934$ &
$-.157$ \\
& $\tilde{\chi}^0_5$ & $-423$ & $0.008$ & $0.231$ & $-.923$ & $-.303$ &
$0.050$ \\ \hline & \multicolumn{7}{|c||}{} \\
Charginos & \multicolumn{7}{|c||}{$m_{\tilde{\chi}^\pm_1} = -50$ GeV
\hspace{1cm}  $m_{\tilde{\chi}^\pm_2}= 418$ GeV } \\
& \multicolumn{7}{|c||}{}   \\ \hline \hline
\multicolumn{8}{||c||}{} \\
\multicolumn{8}{||c||}{Scenario B} \\
\multicolumn{8}{||c||}{$M=120$ GeV, $x=1000$ GeV, $\lambda=0.4$, $k=0.001$,
$\tan\beta=2$} \\ \multicolumn{8}{||c||}{} \\ \hline
& & Mass & Photino & Zino & Higgsino A & Higgsino B & Singlet \\
& & [GeV] & & & & & \\
& $\tilde{\chi}^0_1$ & $-10$ & $-.054$ & $0.135$ & $0.127$ & $-.113$ &
$0.975$ \\
& $\tilde{\chi}^0_2$ & $-55$ & $0.784$ & $-.589$ & $-.115$ & $-.097$ &
$0.129$ \\
Neutralinos & $\tilde{\chi}^0_3$ & $-104$ & $-.618$ & $-.750$ &
$-.186$ & $-.118$ & $0.080$ \\
& $\tilde{\chi}^0_4$ & $412$ & $-.003$ & $0.055$ & $0.313$ & $-.935$ &
$-.157$ \\
& $\tilde{\chi}^0_5$ & $-425$ & $0.019$ & $0.264$ & $-.915$ & $-.230$ &
$0.049$ \\ \hline & \multicolumn{7}{|c||}{} \\
Charginos & \multicolumn{7}{|c||}{$m_{\tilde{\chi}^\pm_1} = -102$ GeV
\hspace{1cm}  $m_{\tilde{\chi}^\pm_2}= 420$ GeV } \\
& \multicolumn{7}{|c||}{}   \\ \hline \hline
\end{tabular}
\end{center}
\caption{Scenarios A and B with a light singlet-like neutralino
and different gaugino mass parameters $M$.
Shown are the neutralino and chargino masses
and the elements $N_{ij}$ of the neutralino mixing matrix.}
\label{szetabab}
\end{table}
\newpage
\begin{table}
\begin{center}
\begin{tabular}{||l|l|cccccc||}
\hline
\hline
\multicolumn{8}{||c||}{} \\
\multicolumn{8}{||c||}{Scenario C} \\
\multicolumn{8}{||c||}{$M=200$ GeV, $x=1000$ GeV, $\lambda=0.4$, $k=0.001$,
$\tan\beta=2$} \\ \multicolumn{8}{||c||}{} \\ \hline
& & Mass & Photino & Zino & Higgsino A & Higgsino B & Singlet \\
& & [GeV] & & & & & \\
& $\tilde{\chi}^0_1$ & $-11$ & $0.026$ & $-.070$ & $-.116$ & $0.123$ &
$0.983$ \\
& $\tilde{\chi}^0_2$ & $-94$ & $0.812$ & $-.558$ & $-.132$ & $-.088$ &
$0.066$ \\
Neutralinos & $\tilde{\chi}^0_3$ & $-178$ & $-.581$ & $-.757$ &
$-.261$ & $0-133$ & $0.053$ \\
& $\tilde{\chi}^0_4$ & $412$ & $0.004$ & $-.049$ & $-.313$ & $0.935$ &
$0.157$ \\
& $\tilde{\chi}^0_5$ & $-431$ & $0.045$ & $0.329$ & $-.896$ & $-.291$ &
$0.047$ \\ \hline & \multicolumn{7}{|c||}{} \\
Charginos & \multicolumn{7}{|c||}{$m_{\tilde{\chi}^\pm_1} = -176$ GeV
\hspace{1cm}  $m_{\tilde{\chi}^\pm_2}= 427$ GeV } \\
& \multicolumn{7}{|c||}{}
\\ \hline \hline
\multicolumn{8}{||c||}{} \\
\multicolumn{8}{||c||}{Scenario D} \\
\multicolumn{8}{||c||}{$M=-120$ GeV, $x=1000$ GeV, $\lambda=0.4$, $k=0.001$,
$\tan\beta=2$} \\ \multicolumn{8}{||c||}{} \\ \hline
& & Mass & Photino & Zino & Higgsino A & Higgsino B & Singlet \\
& & [GeV] & & & & & \\
& $\tilde{\chi}^0_1$ & $-12$ & $0.022$ & $-.073$ & $0.089$ & $ -.142$ &
$0.983$ \\
& $\tilde{\chi}^0_2$ & $63$ & $0.913$ & $-.396$ & $-.065$ & $-.047$ &
$-.050$ \\
Neutralinos & $\tilde{\chi}^0_3$ & $130$ & $-.407$ & $-.900$ &
$-.116$ & $-.123$ & $-.065$ \\
& $\tilde{\chi}^0_4$ & $413$ & $-.007$ & $-.095$ & $-.316$ & $0.931$ &
$0.156$ \\
& $\tilde{\chi}^0_5$ & $-416$ & $-.008$ & $0.164$ & $-.935$ & $-.310$ &
$0.052$ \\ \hline & \multicolumn{7}{|c||}{} \\
Charginos & \multicolumn{7}{|c||}{$m_{\tilde{\chi}^\pm_1} = 129$ GeV
\hspace{1cm}  $m_{\tilde{\chi}^\pm_2}= 413$ GeV } \\
& \multicolumn{7}{|c||}{}   \\ \hline \hline
\end{tabular}
\end{center}
\caption{Scenarios C and D with a light singlet-like neutralino
and gaugino mass parameters $M$ with different signs.
Shown are the neutralino and chargino masses
and the elements $N_{ij}$ of the neutralino mixing matrix.}
\label{szetabcd}
\end{table}
\newpage
\begin{table}
\begin{center}
\begin{tabular}{||l|l|cccc||}
\hline
\hline
\multicolumn{6}{||c||}{} \\
\multicolumn{6}{||c||}{Higgs bosons} \\
\multicolumn{6}{||c||}{$x=1000$ GeV, $\lambda=0.4$, $k=0.001$,
$\tan\beta=2$, $A_t=0$ GeV} \\
\multicolumn{6}{||c||}{ $m_{\tilde{t}_1}=150$ GeV,
$m_{\tilde{t}_2}=500$ GeV}  \\ \multicolumn{6}{||c||}{} \\ \hline
& & Mass & $H_1^0$& $H_2^0$& $N$ \\
& & [GeV] & & &  \\
& $S_1$ & $ 37-54$ & $0.001-0.114$ & $ 0-0.318$ & $0.941- 0.999 $  \\
Scalars & $S_2$ & $ 96-100$ & $0.436-0.451$ & $ 0.835-0.894$ & $0.004-0.335$ \\
& $S_3$ & $ 972-1006$ & $0.893$ & $ 0.449$ & $ 0.043 $ \\
& & \multicolumn{4}{|c||}{} \\ \hline & & \multicolumn{4}{|c||}{} \\
Pseudoscalars & $P_1$ & $ 7-51$ & $0.062$ & $ 0.031$ &$ 0.998$\\
& $P_2$ & $ 973-1006$ & $0.892$ & $ 0.446$ & $0.069$ \\
& & \multicolumn{4}{|c||}{} \\ \hline \hline
\end{tabular}
\end{center}
\caption{Higgs masses $m_{S_a}$, $m_{P_\alpha}$ and mixings $U^S_{ab}$,
$U^P_{\alpha \beta}$, respectively, in the scenarios A -- D.}
\label{szetabhiggsa}
\end{table}
\newpage
\begin{table}
\begin{center}
\begin{tabular}{||l|l|cccccc||}
\hline
\hline
\multicolumn{8}{||c||}{} \\
\multicolumn{8}{||c||}{Scenario E} \\
\multicolumn{8}{||c||}{$M=115$ GeV, $x=1000$ GeV, $\lambda=0.4$, $k=0.035$,
$\tan\beta=2$} \\ \multicolumn{8}{||c||}{} \\ \hline
& & Mass & Photino & Zino & Higgsino A & Higgsino B & Singlet \\
& & [GeV] & & & & & \\
& $\tilde{\chi}^0_1$ & $-50$ & $-.735$ & $0.621$ & $0.156$ & $0.061$ &
$0.214$ \\
& $\tilde{\chi}^0_2$ & $-77$ & $0.360$ & $0.089$ & $0.120$ & $-.088$ &
$0.917$ \\
Neutralinos & $\tilde{\chi}^0_3$ & $-102$ & $-.574$ & $-.732$ &
$-.154$ & $-.138$ & $0.303$ \\
& $\tilde{\chi}^0_4$ & $411$ & $-.003$ & $0.056$ & $0.315$ & $-.938$ &
$-.136$ \\
& $\tilde{\chi}^0_5$ & $-425$ & $0.018$ & $0.260$ & $-.916$ & $-.301$ &
$0.059$ \\ \hline & \multicolumn{7}{|c||}{} \\
Charginos & \multicolumn{7}{|c||}{$m_{\tilde{\chi}^\pm_1} = -97$ GeV
\hspace{1cm}  $m_{\tilde{\chi}^\pm_2}= 420$ GeV } \\
& \multicolumn{7}{|c||}{}   \\ \hline
\hline
\multicolumn{8}{||c||}{} \\
\multicolumn{8}{||c||}{Scenario F} \\
\multicolumn{8}{||c||}{$M=-95$ GeV, $x=1000$ GeV, $\lambda=0.4$, $k=0.035$,
$\tan\beta=2$} \\ \multicolumn{8}{||c||}{} \\ \hline
& & Mass & Photino & Zino & Higgsino A & Higgsino B & Singlet \\
& & [GeV] & & & & & \\
& $\tilde{\chi}^0_1$ & $50$ & $0.922$ & $-.379$ & $-.062$ & $-.048$ &
$-.028$ \\
& $\tilde{\chi}^0_2$ & $-78$ & $0.008$ & $-.054$ & $0.094$ & $-.116$ &
$0.988$ \\
Neutralinos & $\tilde{\chi}^0_3$ & $105$ & $-.387$ & $0.904$ &
$-.124$ & $-.123$ & $-.049$ \\
& $\tilde{\chi}^0_4$ & $412$ & $-.005$ & $-.089$ & $-.318$ & $0.934$ &
$0.135$ \\
& $\tilde{\chi}^0_5$ & $-416$ & $0.008$ & $-.170$ & $0.933$ & $0.310$ &
$-.062$ \\ \hline & \multicolumn{7}{|c||}{} \\
Charginos & \multicolumn{7}{|c||}{$m_{\tilde{\chi}^\pm_1} = 104$ GeV
\hspace{1cm}  $m_{\tilde{\chi}^\pm_2}= 413$ GeV } \\
& \multicolumn{7}{|c||}{}   \\ \hline
& \multicolumn{7}{|c||}{} \\
& \multicolumn{7}{|c||}{$A_t=0$ GeV, $m_{\tilde{t}_1}=150$ GeV,
$m_{\tilde{t}_2}=500$ GeV} \\ & \multicolumn{7}{|c||}{} \\
Higgs bosons & \multicolumn{7}{|c||}{$m_{S_1}=43-86$ GeV,
$m_{S_2}=96-119$ GeV,
$m_{S_3}=930-1028$ GeV} \\ & \multicolumn{7}{|c||}{$m_{P_1}=38-123$ GeV,
$m_{P_2}=930-1029$ GeV} \\ & \multicolumn{7}{|c||}{}   \\ \hline \hline
\end{tabular}
\end{center}
\caption{Scenarios E und F with a singlet-like second lightest neutralino.
Shown are the neutralino and chargino masses
and the elements $N_{ij}$ of the neutralino mixing matrix
as well as the range of the Higgs masses.}
\label{szetabef}
\end{table}
\newpage
\begin{table}
\begin{center}
\begin{tabular}{||l|l|cccccc||}
\hline \hline
\multicolumn{8}{||c||}{} \\
\multicolumn{8}{||c||}{Scenario G} \\
\multicolumn{8}{||c||}{$M=-90$ GeV, $x=300$ GeV, $\lambda=0.4$, $k=0.1$,
$\tan\beta=2$} \\ \multicolumn{8}{||c||}{} \\ \hline
& & Mass & Photino & Zino & Higgsino A & Higgsino B & Singlet \\
& & [GeV] & & & & & \\
& $\tilde{\chi}^0_1$ & $50$ & $0.953$ & $-.258$ & $-.115$ & $-.092$ &
$-.058$ \\
& $\tilde{\chi}^0_2$ & $-79$ & $0.022$ & $-.154$ & $0.264$ & $-.256$ &
$0.917$ \\
Neutralinos & $\tilde{\chi}^0_3$ & $105$ & $-.286$ & $-.750$ &
$-.150$ & $-.532$ & $-.224$ \\
& $\tilde{\chi}^0_4$ & $156$ & $-.091$ & $0.485$ & $0.387$ & $-.741$ &
$-.238$ \\
& $\tilde{\chi}^0_5$ & $-157$ & $-.030$ & $0.335$ & $-.863$ & $-.306$ &
$0.220$ \\ \hline & \multicolumn{7}{|c||}{} \\
Charginos & \multicolumn{7}{|c||}{$m_{\tilde{\chi}^\pm_1} = 100$ GeV
\hspace{1cm}  $m_{\tilde{\chi}^\pm_2}= 159$ GeV } \\
& \multicolumn{7}{|c||}{}   \\ \hline
& \multicolumn{7}{|c||}{} \\
& \multicolumn{7}{|c||}{$A_t=0$ GeV, $m_{\tilde{t}_1}=150$ GeV,
$m_{\tilde{t}_2}=500$ GeV} \\ & \multicolumn{7}{|c||}{} \\
Higgs bosons & \multicolumn{7}{|c||}{$m_{S_1}=7-73$ GeV,
$m_{S_2}=91-107$ GeV,
$m_{S_3}=252-326$ GeV} \\ & \multicolumn{7}{|c||}{$m_{P_1}=60-136$ GeV,
$m_{P_2}=245-323$ GeV} \\ & \multicolumn{7}{|c||}{}   \\ \hline \hline
\end{tabular}
\end{center}
\caption{Scenario G with a smaller singlet vacuum expectation value $x$
and a singlet-like second lightest neutralino.
Shown are the neutralino and chargino masses
and the elements $N_{ij}$ of the neutralino mixing matrix
as well as the range of the Higgs masses.}
\label{szetabg}
\end{table}
\newpage
\begin{table}
\begin{center}
\begin{tabular}{||l||c|c||c|c||c|c||}
\hline \hline & \multicolumn{2}{c||}{} & \multicolumn{2}{c||}{} &
\multicolumn{2}{c||}{} \\
& \multicolumn{2}{c||}{Scenario E} & \multicolumn{2}{c||}{Scenario F} &
\multicolumn{2}{c||}{Scenario G}  \\
& \multicolumn{2}{c||}{} & \multicolumn{2}{c||}{} &
\multicolumn{2}{c||}{} \\
& \multicolumn{2}{c||}{$A_\lambda=900$ GeV} &
\multicolumn{2}{c||}{$A_\lambda=900$ GeV} &
\multicolumn{2}{c||}{$A_\lambda=250$ GeV} \\
& \multicolumn{2}{c||}{$A_k=50$ GeV} &
\multicolumn{2}{c||}{$A_k=50$ GeV} &
\multicolumn{2}{c||}{$A_k=30$ GeV} \\
& \multicolumn{6}{c||}{$A_t=0$ GeV, $m_{\tilde{t}_1}=150$ GeV,
$m_{\tilde{t}_2}=500$ GeV} \\
& \multicolumn{2}{c||}{} & \multicolumn{2}{c||}{} &
\multicolumn{2}{c||}{} \\
\cline{2-7} \cline{2-7}
& $\Gamma /$GeV & BR & $\Gamma /$GeV & BR & $\Gamma /$GeV & BR \\ \hline
$\tilde{\chi}^0_2 \longrightarrow \tilde{\chi}^0_1 e^+e^- $ &
$8.60 \cdot 10^{-9} $  &$0.55$  &$9.60\cdot 10^{-11}$  & $0.07$ &
$1.29 \cdot 10^{-9} $& $0.14$ \\ \hline
$\tilde{\chi}^0_2 \longrightarrow \tilde{\chi}^0_1 \nu \bar{\nu} $ &
$1.82 \cdot 10^{-9} $&$0.12$  &$1.47  \cdot 10^{-11}$ & $ 0.01$ &
$7.25 \cdot 10^{-10} $& $0.08$ \\ \hline
$\tilde{\chi}^0_2 \longrightarrow \tilde{\chi}^0_1 q \bar{q} $ &
$3.23 \cdot 10^{-9} $ &$0.21$  &$3.23 \cdot 10^{-10}$ &$0.22$ &
$5.80 \cdot 10^{-9} $ & $0.61$ \\ \hline
$\tilde{\chi}^0_2 \longrightarrow \tilde{\chi}^0_1 \gamma $ &
$1.83 \cdot 10^{-9} $ &$0.12$  &$1.00 \cdot 10^{-9} $ &$0.70$ &
$1.60 \cdot 10^{-9} $ & $0.17$ \\
\hline \hline
$\tilde{\chi}^0_3\longrightarrow \tilde{\chi}^0_1 e^+e^- $ &
$9.95 \cdot 10^{-8}$ & $0.10$ & $5.21 \cdot 10^{-7}$ & $0.39$ &
$3.14 \cdot 10^{-7}$ & $0.10$ \\ \hline
$\tilde{\chi}^0_3 \longrightarrow \tilde{\chi}^0_1 \nu \bar{\nu} $ &
$2.69 \cdot 10^{-7}$ & $0.26$ & $2.19 \cdot 10^{-7}$ & $0.17$ &
$1.05 \cdot 10^{-7}$ & $0.03$ \\ \hline
$\tilde{\chi}^0_3 \longrightarrow \tilde{\chi}^0_1 q \bar{q} $ &
$5.86 \cdot 10^{-7}$ & $0.57$ & $5.48 \cdot 10^{-7}$ & $0.41$ &
$2.75 \cdot 10^{-6}$ & $0.86$ \\ \hline
$\tilde{\chi}^0_3 \longrightarrow \tilde{\chi}^0_1 \gamma $ &
$6.06 \cdot 10^{-9}$ & $0.006$ & $9.88 \cdot 10^{-9}$ & $0.007$ &
$2.16 \cdot 10^{-8}$ & $0.007$ \\ \hline
$\tilde{\chi}^0_3 \longrightarrow \tilde{\chi}^\pm_1 e^\mp \nu$ &
$1.64 \cdot 10^{-9}$ & $0.001$ & $7.08 \cdot 10^{-13}$ & $<0.001$ &
$2.47 \cdot 10^{-9}$ & $<0.001$ \\ \hline
$\tilde{\chi}^0_3 \longrightarrow \tilde{\chi}^\pm_1 qq'$ &
$4.92 \cdot 10^{-9}$ & $0.005$ & $2.12 \cdot 10^{-12}$ & $<0.001$ &
$7.39 \cdot 10^{-9}$ & $<0.001$ \\
\hline \hline
$\tilde{\chi}^0_3\longrightarrow \tilde{\chi}^0_2 e^+e^- $ &
$7.24 \cdot 10^{-9}$ & $0.007$ & $2.16 \cdot 10^{-10}$ & $<0.001$ &
$4.38 \cdot 10^{-10}$ & $<0.001$ \\ \hline
$\tilde{\chi}^0_3 \longrightarrow \tilde{\chi}^0_2 \nu \bar{\nu} $ &
$1.62 \cdot 10^{-9}$ & $0.001$ & $2.08 \cdot 10^{-11}$ & $<0.001$ &
$2.48 \cdot 10^{-11}$ & $<0.001$ \\ \hline
$\tilde{\chi}^0_3 \longrightarrow \tilde{\chi}^0_2 q \bar{q} $ &
$2.24  \cdot 10^{-8}$ & $0.02$ & $1.54 \cdot 10^{-9}$ & $0.001$&
$3.41 \cdot 10^{-9}$ & $<0.001$ \\ \hline
$\tilde{\chi}^0_3 \longrightarrow \tilde{\chi}^0_2 \gamma $ &
$2.59 \cdot 10^{-8}$ & $0.02$ & $2.75 \cdot 10^{-8}$ & $0.02$ &
$2.16 \cdot 10^{-8}$ & $0.007$ \\ \hline  \hline
\end{tabular}
\end{center}
\caption{Decay widths and branching ratios for the decays of the second
and third neutralinos in the scenarios E -- G.}
\label{tabefg}
\end{table}
\newpage
%%%%%%%%%% Figure 1 %%%%%%%%%%%%%%%%%%%%%%%%%%%%%
\newpage
%%%%%%%%%% Figure 2 %%%%%%%%%%%%%%%%%%%%%%%%%%%%%
\begin{figure}
\begin{center}
\begin{picture}(16,5)
\put(4,0){\includegraphics{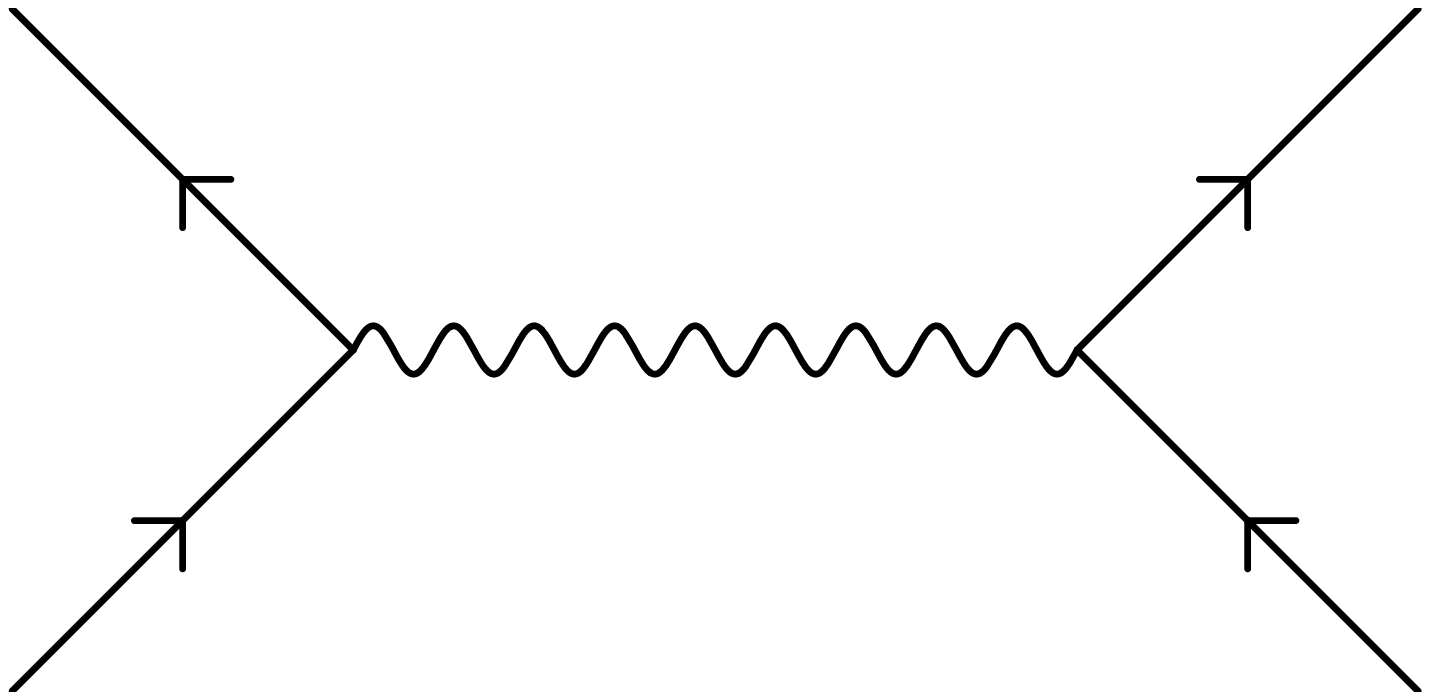}}
\put(5.4,4.2){\line(1,0){0.7}}
\put(9,4.2){\line(1,0){0.7}}
\put(4.4,2.0){\line(3,-1){1.5}}
\put(4.4,0.4){\line(3,1){1.5}}
\put(11,2.0){\line(-3,-1){1.5}}
\put(11,0.4){\line(-3,1){1.5}}
\put(1,4){$\frac{ig}{\cos\theta_W}\gamma^\mu (L_{e}P_L+
R_{e}P_R$)}
\put(10,4){$\frac{ig}{\cos\theta_W}\gamma^\mu (O_{ij}^{''L}P_L+
O_{ij}^{''R}P_R$)}
\put(6.3,1.1){$gf_{ei}^LP_R+gf_{ei}^RP_L$}
\put(1,-3){\includegraphics{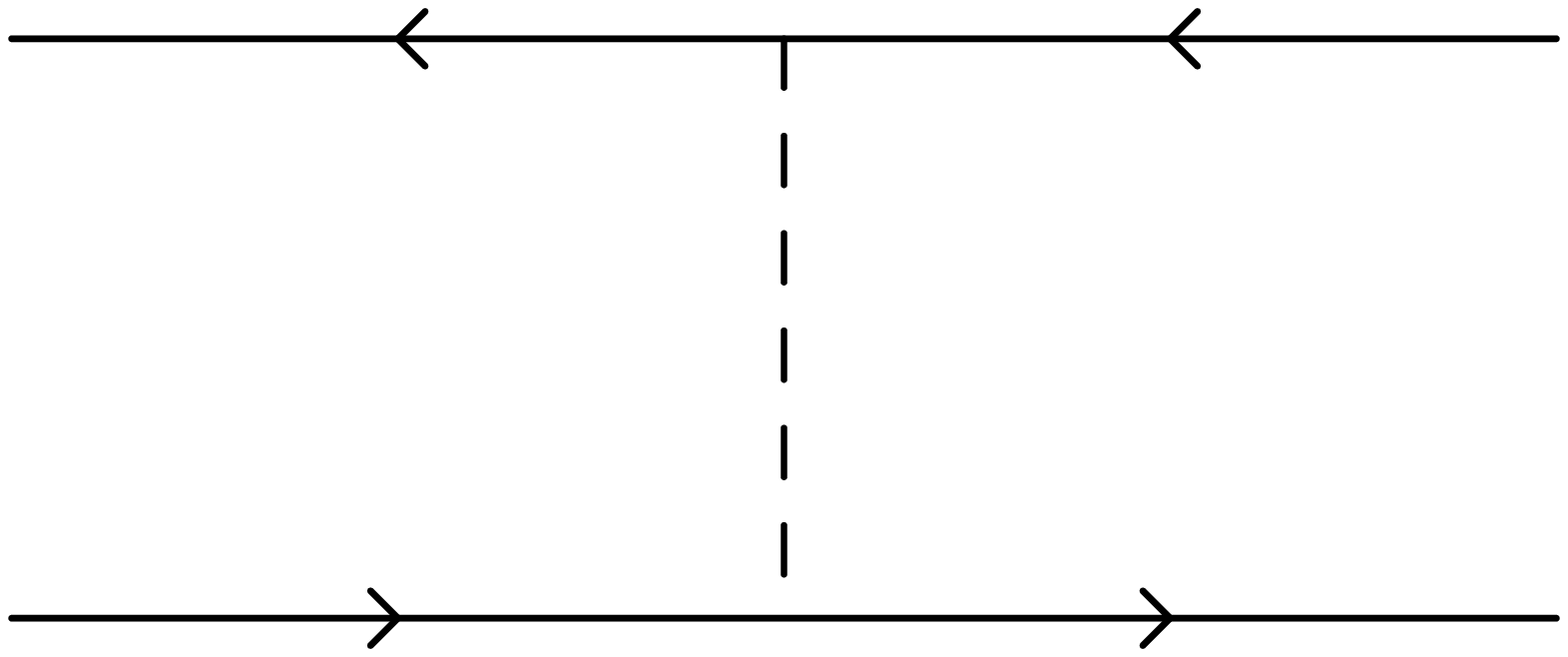}}
\put(8,-3){\includegraphics{nfig2b.ps}}
\put(4.50,5.3){$e^+$}
\put(4.50,3){$e^-$}
\put(10,5.3){$\tilde{\chi}_i^0$}
\put(10,3){$\tilde{\chi}_j^0$}
\put(7.10,3.7){$Z$}
\put(1.50,2.3){$e^+$}
\put(1.50,0){$e^-$}
\put(7,2.1){$\tilde{\chi}_i^0$}
\put(7,0.1){$\tilde{\chi}_j^0$}
\put(2.85,0.9){$\tilde{e}_L$, $\tilde{e}_R$}
\put(8.50,2.3){$e^+$}
\put(8.50,0){$e^-$}
\put(14,2.1){$\tilde{\chi}_j^0$}
\put(14,0.1){$\tilde{\chi}_i^0$}
\put(11.35,0.9){$\tilde{e}_L$, $\tilde{e}_R$}
\end{picture}
\end{center}
\caption{Feynman graphs and vertex factors for the process
$e^+e^- \longrightarrow \tilde{\chi}_i^0
\tilde{\chi}_j^0$ ($i,j=1,\ldots,5$). Explicit expressions for the
parameters $L_e$, $R_e$, $f_{ei}^L$, $f_{ei}^R$, $O_{ij}^{''L}$,
$O_{ij}^{''R}$ are given in the appendix.}
\label{graph}
\end{figure}
\newpage
%%%%%%%%%% Figure 3 %%%%%%%%%%%%%%%%%%%%%%%%%%%%%
\newpage
%%%%%%%%%% Figure 4 %%%%%%%%%%%%%%%%%%%%%%%%%%%%%
\begin{figure}
\begin{center}
\begin{picture}(16,6.0)
\put(0,0){\includegraphics{nfig4.ps}}
\put(5.0,-0.5){$\sqrt{s}/$GeV}
\put(4.0,0.7){Scenario D}
\put(-0.0,6.0){$\sigma/$pb}
\put(9,-0.5){\includegraphics{beschr.ps}}
\put(12.5,6.3){$e^+e^- \rightarrow \tilde{\chi}^0_1 \tilde{\chi}^0_2$}
\put(12.5,5.5){$e^+e^- \rightarrow \tilde{\chi}^0_1 \tilde{\chi}^0_3$}
\put(12.5,4.7){$e^+e^- \rightarrow \tilde{\chi}^0_2 \tilde{\chi}^0_2$}
\put(12.5,4.0){$e^+e^- \rightarrow \tilde{\chi}^0_2 \tilde{\chi}^0_3$}
\put(12.5,3.2){$e^+e^- \rightarrow \tilde{\chi}^0_3 \tilde{\chi}^0_3$}
\put(12.5,2.4){$e^+e^- \rightarrow \tilde{\chi}^0_1 \tilde{\chi}^0_4$}
\put(12.5,1.7){$e^+e^- \rightarrow \tilde{\chi}^0_2 \tilde{\chi}^0_4$}
\put(12.5,0.9){$e^+e^- \rightarrow \tilde{\chi}^0_1 \tilde{\chi}^0_5$}
\put(12.5,0.1){$e^+e^- \rightarrow \tilde{\chi}^0_2 \tilde{\chi}^0_5$}
\end{picture}
\end{center}
\caption{Cross sections for
$e^+e^- \longrightarrow \tilde{\chi}^0_i \tilde{\chi}^0_j$
in scenario D.}
\label{prodfigd}
\end{figure}
\newpage
%%%%%%%%%% Figure 5 %%%%%%%%%%%%%%%%%%%%%%%%%%%%%
\newpage
%%%%%%%%%% Figure 6 %%%%%%%%%%%%%%%%%%%%%%%%%%%%%
\newpage
%%%%%%%%%% Figure 7 %%%%%%%%%%%%%%%%%%%%%%%%%%%%%
\newpage
%%%%%%%%%% Figure 8 %%%%%%%%%%%%%%%%%%%%%%%%%%%%%
\newpage
%%%%%%%%%% Figure 9 %%%%%%%%%%%%%%%%%%%%%%%%%%%%%
\newpage
%%%%%%%%%% Figure 10%%%%%%%%%%%%%%%%%%%%%%%%%%%%%
\end{document}